\documentclass[aps,prd,showpacs,twocolumn,superscriptaddress,groupedaddress,floatfix,amsfonts,nofootinbib]{revtex4}
\usepackage{graphicx}
\usepackage{dcolumn}
\usepackage{bm}
\usepackage{amssymb}
\usepackage{amsmath}
\usepackage{xspace}
\usepackage{color}

\newcommand{\Ptg}{p_{T}^{\gamma}}

\newcommand{\la}{\langle}
\newcommand{\ra}{\rangle}

\newcommand{\gt}{\!>\!}

\newcommand{\gpj}{$\gamma+{\rm jet}$\xspace}
\newcommand{\gpONEj}{$\gamma+{\rm 1~jet}$\xspace}
\newcommand{\gpLEADj}{$\gamma+{\rm leading~ jet}$\xspace}
\newcommand{\gpTWOj}{$\gamma+{\rm 2~jet}$\xspace}
\newcommand{\gpTHRj}{$\gamma+{\rm 3~jet}$\xspace}
\newcommand{\gpTHRjX}{$\gamma+{\rm 3~jet}+X$\xspace}
\newcommand{\TWOjj}{${\rm jet2}+{\rm jet3}$\xspace}

\newcommand{\ptsj}{p_T^{\rm jet 2}}
\newcommand{\dS}{$\Delta S$~}
\newcommand{\dPhi}{$\Delta \phi$~}

\newcommand{\DPhi}{\Delta\phi}

\newcommand{\DSigDPhi}{$(1/\sigma_{\gamma 2j}) d\sigma_{\gamma 2j}/d\Delta\phi$\xspace}
\newcommand{\DSigDS}{$(1/\sigma_{\gamma 3j})d\sigma_{\gamma 3j}/d\Delta S$\xspace}

\newcommand{\CHECK}[1]{\textbf{\color{red}[#1]}\xspace}
\newcommand{\chk}[1]{\CHECK{CHECK THIS!}}

\begin{document}
\widetext

\hspace{5.2in}\mbox{ FERMILAB-PUB-11-007-E}

\title{\boldmath Azimuthal decorrelations and multiple parton
  interactions in \gpTWOj and \gpTHRj events in $p\bar{p}$ collisions 
at $\sqrt{s}=1.96$~TeV}

%
\affiliation{Universidad de Buenos Aires, Buenos Aires, Argentina}
\affiliation{LAFEX, Centro Brasileiro de Pesquisas F{\'\i}sicas, Rio de Janeiro, Brazil}
\affiliation{Universidade do Estado do Rio de Janeiro, Rio de Janeiro, Brazil}
\affiliation{Universidade Federal do ABC, Santo Andr\'e, Brazil}
\affiliation{Instituto de F\'{\i}sica Te\'orica, Universidade Estadual Paulista, S\~ao Paulo, Brazil}
\affiliation{Simon Fraser University, Vancouver, British Columbia, and York University, Toronto, Ontario, Canada}
\affiliation{University of Science and Technology of China, Hefei, People's Republic of China}
\affiliation{Universidad de los Andes, Bogot\'{a}, Colombia}
\affiliation{Charles University, Faculty of Mathematics and Physics, Center for Particle Physics, Prague, Czech Republic}
\affiliation{Czech Technical University in Prague, Prague, Czech Republic}
\affiliation{Center for Particle Physics, Institute of Physics, Academy of Sciences of the Czech Republic, Prague, Czech Republic}
\affiliation{Universidad San Francisco de Quito, Quito, Ecuador}
\affiliation{LPC, Universit\'e Blaise Pascal, CNRS/IN2P3, Clermont, France}
\affiliation{LPSC, Universit\'e Joseph Fourier Grenoble 1, CNRS/IN2P3, Institut National Polytechnique de Grenoble, Grenoble, France}
\affiliation{CPPM, Aix-Marseille Universit\'e, CNRS/IN2P3, Marseille, France}
\affiliation{LAL, Universit\'e Paris-Sud, CNRS/IN2P3, Orsay, France}
\affiliation{LPNHE, Universit\'es Paris VI and VII, CNRS/IN2P3, Paris, France}
\affiliation{CEA, Irfu, SPP, Saclay, France}
\affiliation{IPHC, Universit\'e de Strasbourg, CNRS/IN2P3, Strasbourg, France}
\affiliation{IPNL, Universit\'e Lyon 1, CNRS/IN2P3, Villeurbanne, France and Universit\'e de Lyon, Lyon, France}
\affiliation{III. Physikalisches Institut A, RWTH Aachen University, Aachen, Germany}
\affiliation{Physikalisches Institut, Universit{\"a}t Freiburg, Freiburg, Germany}
\affiliation{II. Physikalisches Institut, Georg-August-Universit{\"a}t G\"ottingen, G\"ottingen, Germany}
\affiliation{Institut f{\"u}r Physik, Universit{\"a}t Mainz, Mainz, Germany}
\affiliation{Ludwig-Maximilians-Universit{\"a}t M{\"u}nchen, M{\"u}nchen, Germany}
\affiliation{Fachbereich Physik, Bergische Universit{\"a}t Wuppertal, Wuppertal, Germany}
\affiliation{Panjab University, Chandigarh, India}
\affiliation{Delhi University, Delhi, India}
\affiliation{Tata Institute of Fundamental Research, Mumbai, India}
\affiliation{University College Dublin, Dublin, Ireland}
\affiliation{Korea Detector Laboratory, Korea University, Seoul, Korea}
\affiliation{CINVESTAV, Mexico City, Mexico}
\affiliation{FOM-Institute NIKHEF and University of Amsterdam/NIKHEF, Amsterdam, The Netherlands}
\affiliation{Radboud University Nijmegen/NIKHEF, Nijmegen, The Netherlands}
\affiliation{Joint Institute for Nuclear Research, Dubna, Russia}
\affiliation{Institute for Theoretical and Experimental Physics, Moscow, Russia}
\affiliation{Moscow State University, Moscow, Russia}
\affiliation{Institute for High Energy Physics, Protvino, Russia}
\affiliation{Petersburg Nuclear Physics Institute, St. Petersburg, Russia}
\affiliation{Stockholm University, Stockholm and Uppsala University, Uppsala, Sweden }
\affiliation{Lancaster University, Lancaster LA1 4YB, United Kingdom}
\affiliation{Imperial College London, London SW7 2AZ, United Kingdom}
\affiliation{The University of Manchester, Manchester M13 9PL, United Kingdom}
\affiliation{University of Arizona, Tucson, Arizona 85721, USA}
\affiliation{University of California Riverside, Riverside, California 92521, USA}
\affiliation{Florida State University, Tallahassee, Florida 32306, USA}
\affiliation{Fermi National Accelerator Laboratory, Batavia, Illinois 60510, USA}
\affiliation{University of Illinois at Chicago, Chicago, Illinois 60607, USA}
\affiliation{Northern Illinois University, DeKalb, Illinois 60115, USA}
\affiliation{Northwestern University, Evanston, Illinois 60208, USA}
\affiliation{Indiana University, Bloomington, Indiana 47405, USA}
\affiliation{Purdue University Calumet, Hammond, Indiana 46323, USA}
\affiliation{University of Notre Dame, Notre Dame, Indiana 46556, USA}
\affiliation{Iowa State University, Ames, Iowa 50011, USA}
\affiliation{University of Kansas, Lawrence, Kansas 66045, USA}
\affiliation{Kansas State University, Manhattan, Kansas 66506, USA}
\affiliation{Louisiana Tech University, Ruston, Louisiana 71272, USA}
\affiliation{Boston University, Boston, Massachusetts 02215, USA}
\affiliation{Northeastern University, Boston, Massachusetts 02115, USA}
\affiliation{University of Michigan, Ann Arbor, Michigan 48109, USA}
\affiliation{Michigan State University, East Lansing, Michigan 48824, USA}
\affiliation{University of Mississippi, University, Mississippi 38677, USA}
\affiliation{University of Nebraska, Lincoln, Nebraska 68588, USA}
\affiliation{Rutgers University, Piscataway, New Jersey 08855, USA}
\affiliation{Princeton University, Princeton, New Jersey 08544, USA}
\affiliation{State University of New York, Buffalo, New York 14260, USA}
\affiliation{Columbia University, New York, New York 10027, USA}
\affiliation{University of Rochester, Rochester, New York 14627, USA}
\affiliation{State University of New York, Stony Brook, New York 11794, USA}
\affiliation{Brookhaven National Laboratory, Upton, New York 11973, USA}
\affiliation{Langston University, Langston, Oklahoma 73050, USA}
\affiliation{University of Oklahoma, Norman, Oklahoma 73019, USA}
\affiliation{Oklahoma State University, Stillwater, Oklahoma 74078, USA}
\affiliation{Brown University, Providence, Rhode Island 02912, USA}
\affiliation{University of Texas, Arlington, Texas 76019, USA}
\affiliation{Southern Methodist University, Dallas, Texas 75275, USA}
\affiliation{Rice University, Houston, Texas 77005, USA}
\affiliation{University of Virginia, Charlottesville, Virginia 22901, USA}
\affiliation{University of Washington, Seattle, Washington 98195, USA}
\author{V.M.~Abazov} \affiliation{Joint Institute for Nuclear Research, Dubna, Russia}
\author{B.~Abbott} \affiliation{University of Oklahoma, Norman, Oklahoma 73019, USA}
\author{B.S.~Acharya} \affiliation{Tata Institute of Fundamental Research, Mumbai, India}
\author{M.~Adams} \affiliation{University of Illinois at Chicago, Chicago, Illinois 60607, USA}
\author{T.~Adams} \affiliation{Florida State University, Tallahassee, Florida 32306, USA}
\author{G.D.~Alexeev} \affiliation{Joint Institute for Nuclear Research, Dubna, Russia}
\author{G.~Alkhazov} \affiliation{Petersburg Nuclear Physics Institute, St. Petersburg, Russia}
\author{A.~Alton$^{a}$} \affiliation{University of Michigan, Ann Arbor, Michigan 48109, USA}
\author{G.~Alverson} \affiliation{Northeastern University, Boston, Massachusetts 02115, USA}
\author{G.A.~Alves} \affiliation{LAFEX, Centro Brasileiro de Pesquisas F{\'\i}sicas, Rio de Janeiro, Brazil}
% ...
%
%
\collaboration{The D0 Collaboration\footnote{with visitors from
$^{a}$Augustana College, Sioux Falls, SD, USA,
$^{b}$The University of Liverpool, Liverpool, UK,
$^{c}$SLAC, Menlo Park, CA, USA,
$^{d}$ICREA/IFAE, Barcelona, Spain,
$^{e}$Centro de Investigacion en Computacion - IPN, Mexico City, Mexico,
$^{f}$ECFM, Universidad Autonoma de Sinaloa, Culiac\'an, Mexico,
and 
$^{g}$Universit{\"a}t Bern, Bern, Switzerland.%
}} \noaffiliation
\vskip 0.25cm

\date{January 7, 2011}

\begin{abstract}
     Samples of inclusive \gpTWOj and  \gpTHRj events collected by the D0 experiment 
     with an integrated luminosity of about 1~fb$^{-1}$ in $p\bar{p}$ collisions 
     at  $\sqrt{s}=1.96$ TeV  are used to measure cross sections 
     as a function of the angle in the plane transverse to the beam direction
     between the transverse momentum ($p_T$)
     of the $\gamma+$leading jet system (jets are ordered in $p_T$) 
     and $p_T$ of the other jet for \gpTWOj, 
     or $p_T$ sum of the two other jets for \gpTHRj events.
     The results are compared to different models of multiple parton interactions (MPI)
     in the {\sc pythia} and {\sc sherpa} Monte Carlo (MC) generators. 
     The data indicate a contribution from events with 
     double parton (DP) interactions and are well described by predictions
     provided by the {\sc pythia} MPI models with $p_T$-ordered showers
     and by {\sc sherpa} with the default MPI model.
     The \gpTWOj data are also used to determine the fraction of events
     with DP interactions as a function of 
     the azimuthal angle and as a function of the second jet $p_T$.

\end{abstract}
\maketitle


\section{Introduction}
\label{Sec:Intro}

 The high energy scattering of two nucleons can be considered, in a simplified model, 
 as a single collision of one parton (quark or gluon) 
 from one nucleon with one parton from the other nucleon. 
 In this approach, the remaining ``spectator" partons, which do not take part in the hard
 $2 \to 2$ parton collision, participate in the so-called ``underlying event." 
  However,  there are also models 
  that allow for the possibility that two or more parton pairs undergo a hard interaction
  when two hadrons collide.
  These MPI events have been examined in many theoretical papers 
  \cite{Landsh,Goebel, TH1, TH11, TH2, TH21, TH3, Mang, Sjost, Perugia, PYT, pT_order, Snigir, Snigir1, Trel_05, GS, Snigir2}.
  A comprehensive review of MPI models 
  in hadron collisions is given in \cite{Sjost}.
  A significant amount of experimental data, from the CERN ISR $pp$ collider \cite{AFS}, the
  CERN SPS $p\bar{p}$ collider \cite{UA2}, the Fermilab Tevatron $p\bar{p}$ collider 
  \cite{CDF93, CDF97, E735, D003, D0_2010} and 
  the DESY HERA $ep$ collider \cite{ZEUS,H1}, shows clear evidence for MPI events.

  In addition to parton distribution functions (PDF) and parton cross sections,
  the rates of double and triple parton scattering
  also depend on how the partons are spatially distributed within the hadron. 
  The spatial parton distributions are implemented in 
  various phenomenological models that have been proposed over
  the last $25$ years. They have evolved from the first ``simple''
  model suggested in \cite{TH3}, 
  to more sophisticated models \cite{Perugia,PYT} that 
  consider MPI with correlations in parton momentum and color, as well as effects
  balancing MPI and initial and final state radiation (ISR and FSR) effects,
  which are implemented in the recent (``$p_T$-ordered'') models \cite{pT_order}.

  Beyond the motivation of better understanding non-perturbative quantum chromodynamics (QCD),
  a more realistic modeling of the underlying event and an estimate of the contributions
  from DP interactions are important for  studying 
  background events for many rare processes, including searches for the Higgs boson \cite{WH,WH1,Huss,Berger_dp,DPWH}. 
  Uncertainties in the choice of the underlying event model and related corrections
  also cause uncertainty in measurements of the top quark mass. 
  This uncertainty can be as large as $1.0$ GeV \cite{Skands_top},
  a value obtained from a comparison of MPI models with virtuality-ordered \cite{Sjost} (``old'' models)
  and $p_T$-ordered \cite{pT_order} (``new'' models) parton showers.

In a previous paper \cite{D0_2010}, we have studied the \gpTHRj final state
and extracted the fractions of DP events 
from a comparison of angular distributions in data
with templates obtained from a data-driven DP model.
That paper also presented the effective cross section ($\sigma_{\rm eff}$), 
which characterizes the size of the effective parton-parton interaction  region
and can be used to calculate the DP cross sections for various pairs of 
parton scattering processes.

  In this paper, we extend our previous study 
  by measuring differential cross sections for
  the angle in the plane transverse to the beam direction between the $p_{T}$ vector obtained by pairing
  the photon and the leading (ordered in $p_T$) jet and the $p_{T}$ vector of the other (two) jet(s)
  in $\gamma+{\rm 2(3)~jet}+X$ events (referred to below as ``$\gamma$ + 2(3) jet'' events).
  These cross sections are sensitive to the contributions from jets originating from
  additional parton hard interactions (beyond the dominant one) and can be used to improve existing MPI models,
  and to estimate the fractions of such events \cite{CDF_gj_angles,CDF97,D0_2010}. 
  The cross section measurements are performed at the particle level, which means that
  the jets' four-momenta represent the real energy and direction of the jet of stable particles
  resulting from the hadronization process following the $p \bar p$ interaction \cite{ToolsAndJets}.
  The larger statistics in \gpTWOj events allows to subdivide the cross section measurement
  in bins of the second jet $p_T$ ($\ptsj$). This extension increases the sensitivity to various MPI models.

   In contrast with angular, $p_T$, and multiplicity distributions 
of low $p_T$ tracks traditionally 
used to test MPI models \cite{Sjost,Perugia}, we analyze events with high $p_T$ jets ($p_T>15$~GeV). 
Our approach complements the previous one since the MPI models have not been well tested 
in high $p_T$ regimes, yet this kinematic region is the most important for searches for rare processes 
for which DP events are a potential background \cite{WH,WH1,Huss,Berger_dp,DPWH}.

  This paper is organized as follows.
  In Sec.~\ref{Sec:variables}, we describe the variables used in the analysis and motivate our choice
  of selection criteria.
  In Sec.~\ref{Sec:Sel}, we describe the D0 detector and the identification criteria for  
  photons and jets. 
  In Sec.~\ref{Sec:Models}, we describe the theoretical models used for comparison with data. 
  In Sec.~\ref{Sec:Anal}, we discuss the corrections applied to the data in the cross section
  measurements and the related uncertainties.
  The measured cross sections and comparisons with some model predictions are presented in Sec.~\ref{Sec:xs_data_theory}.
  In Sec.~\ref{Sec:dpfrac}, we extract the fraction of
  DP events in the \gpTWOj final state as a function of $\ptsj$. 
  In Sec.~\ref{Sec:tpfrac}, we estimate the fractions of \gpTHRj events
  occurring due to triple parton scattering, in bins of $\ptsj$. 
  Section \ref{Sec:conclusion} presents our conclusions.

\section{Variables}
\label{Sec:variables}

In this paper, we follow the notation used in our previous analysis \cite{D0_2010}
to distinguish between two classes of events. 
In events of the first class, 
the photon and all jets originate from the same 
single parton-parton interaction (SP) with 
hard gluon bremsstrahlung in the initial or final state. 
In the second class,
at least one of the jets originates from an additional
parton interaction and thus we have at least two parton-parton interactions.

To identify events with two independent parton-parton 
scatterings that produce a \gpTHRj final state, we use an angular distribution
sensitive to the kinematics of the DP events \cite{D0_2010}. 
We define an azimuthal angle between the $p_T$ vector 
of the $\gamma+$leading jet system and the $p_T$ sum of the two other jets:
\begin{eqnarray}
\Delta S \equiv \Delta\phi\left(\vec{P}_{T}^{A}, ~\vec{P}_{T}^{B}\right),
\label{eq:DeltaS_var}
\end{eqnarray}
where $\vec{P}_{T}^{A} = {\vec p}_T^{~\gamma} + {\vec p}_T^{\rm ~jet 1}$ 
and $\vec{P}_{T}^{B} = {\vec p}_T^{\rm ~jet 2} + {\vec p}_T^{\rm ~jet 3}$.
\begin{figure}[h]
~\\[-1mm]
\hspace*{2mm} \includegraphics[scale=0.35]{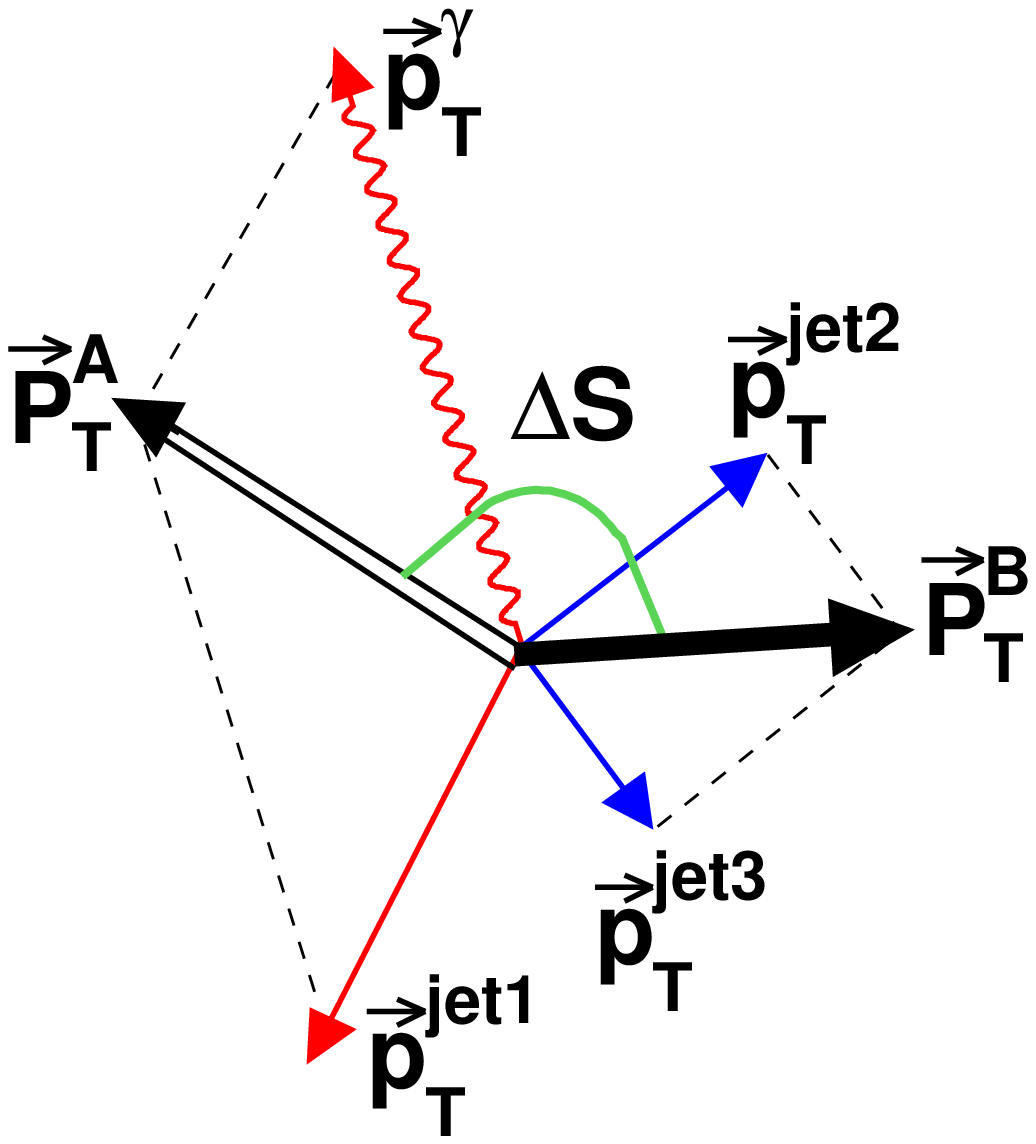}
~\\[-12mm]
\caption{Diagram showing the $p_T$ vectors of the \gpLEADj and \TWOjj systems
in \gpTHRj events.} 
\label{fig:dphi_g3j}
\end{figure}

Figure~\ref{fig:dphi_g3j} shows the sum $p_T$ vectors of the \gpLEADj and \TWOjj systems 
in \gpTHRj events.
In SP events, topologies with two radiated jets emitted close to the leading jet
(recoiling against the photon direction in $\phi$) are preferred
and resulting in a peak at $\Delta S = \pi$.
However, this peak is smeared by the effects of 
additional gluon radiation and detector resolution.
For a simple model of DP events, with both the second and third jets originating from the second parton interaction,
we have exact pairwise balance in $p_T$ in both the \gpj and dijet system,
and thus the $\Delta S$ angle can have any value,
i.e., we expect a uniform $\Delta S$ distribution \cite{Fig9}. 

In this paper, we extend the study of DP interaction to the \gpTWOj events.
In the presence of a DP interaction the second jet in the event originates from a
dijet system in the additional parton interaction and the third jet is
either not reconstructed or below the $p_T$ threshold applied in the event selection.

\begin{figure}[htbp]
~\\[-3mm]
\hspace*{10mm} \includegraphics[scale=0.34]{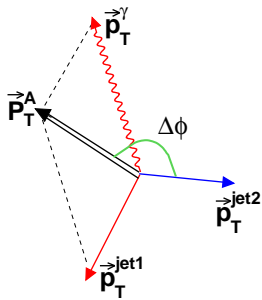}
~\\[-12mm]
\caption{Diagram showing the $p_T$ vectors of the \gpLEADj system and 
$\vec{p}_{T}^{\rm~jet 2}$ in \gpTWOj events.}
\label{fig:dphi_g2j}
\end{figure}

In the case of \gpTWOj events, we introduce a different angular variable, analogous
to (\ref{eq:DeltaS_var}), to retain sensitivity to DP events.
This variable is the azimuthal angle
between the $p_{T}$ vector obtained by pairing the photon and the leading jet $p_T$ vectors 
and the second jet $p_T$ vector:
\begin{eqnarray}
\Delta \phi \equiv \Delta\phi\left(\vec{P}_{T}^{A}, ~\vec{p}_{T}^{\rm ~jet2}\right).
\label{eq:DeltaPHI_var}
\end{eqnarray}
Figure~\ref{fig:dphi_g2j} shows a diagram defining the $p_T$ of the two systems in \gpTWOj events 
and the individual $p_T$ of the objects.
The \dPhi distribution in \gpTWOj events
has been used to estimate the DP fraction by the CDF Collaboration \cite{CDF_gj_angles}.

The $p_T$ spectrum for jets from dijet events falls faster than that for 
jets, resulting from ISR and FSR in \gpj events, 
and thus the DP fractions should depend on the jet $p_T$~\cite{Landsh,TH1,TH11,TH2,TH21,Sjost,D0_2010}.
For this reason, the \dPhi dependent cross sections and the DP fractions in the \gpTWOj events  are measured 
in three $\ptsj$ bins: $15-20$, $20-25$, and $25-30$~GeV.
The \dS dependent cross section is measured  in $\gamma$ + 3 jet events 
(a subsample of the inclusive  $\gamma$ + 2 jet sample) in a single $\ptsj$ interval, $15-30$ GeV.
Such a measurement provides good sensitivity to the DP contribution, and 
discriminating power between different MPI models
because the DP fraction in \gpTHRj events is expected to be higher than 
that in \gpTWOj events. This is expected since the second parton interaction 
will usually produce a dijet final state, 
while the production of an additional jet in SP events via gluon bremsstrahlung
is suppressed by the strong coupling constant $\alpha_s$.

\section{D0 detector and data samples}
\label{Sec:Sel}

The D0 detector is a general purpose detector
described elsewhere in detail~\cite{D0_det}.
Here we briefly describe the detector systems most relevant for this analysis.
Photon candidates are identified 
as isolated clusters of energy deposits in the uranium and liquid-argon sampling calorimeter.
The calorimeter consists of a central section with
coverage in pseudorapidity $|\eta_{\rm det}|<1.1$~\cite{etaphi}
and two end calorimeters covering up to $|\eta_{\rm det}| \approx 4.2$.
The electromagnetic (EM) section of the
calorimeter is segmented longitudinally into four layers, 
with transverse
segmentation into cells of size $\Delta\eta_{\rm det}\times\Delta\phi_{\rm det} = 0.1\times 0.1$,
except for the third layer, where it is $0.05\times 0.05$.
The hadronic portion of the calorimeter is located behind the EM section.
The calorimeter surrounds 
a tracking system consisting of silicon microstrip and
scintillating fiber trackers, both located within a solenoidal magnetic field
of approximately 2~T.

The events used in this analysis are required to pass triggers based on the identification
of high $E_T$ clusters in the EM calorimeter with a shower shape consistent with that expected for photons.
These triggers are $100\%$ efficient for photons with transverse momentum $\Ptg \gt 35$~GeV.
To select photon candidates for our data sample,
we use the following criteria~\cite{gamjet_xs,D0_2010}.
EM objects are reconstructed using a simple cone algorithm with
a cone size ${\cal R}=0.2$ around a seed tower in $\eta-\phi$ space~\cite{gamjet_xs}. 
Regions with poor photon identification capability and limited $\Ptg$ resolution
(found at the boundaries between calorimeter modules and between the central and end
calorimeters) are excluded from the analysis.
Each photon candidate is required to deposit more than 96\% of its detected energy
in the EM section of the calorimeter 
and to be isolated in the annular region between
${\cal R}=0.2$ and ${\cal R}=0.4$ around the 
center of the cluster:
$(E^{\rm iso}_{\rm Tot}-E^{\rm iso}_{\rm Core})/E^{\rm iso}_{\rm Core} < 0.07$, where $E^{\rm iso}_{\rm Tot}$ 
is the total (EM+hadronic) energy in the cone of radius ${\cal R}=0.4$
and $E^{\rm iso}_{\rm Core}$ is the EM tower energy within a radius ${\cal R}=0.2$.
The probability for candidate EM clusters to be spatially matched to 
a reconstructed track is required to be $<0.1\%$,
where  this probability is calculated using the spatial resolutions measured in data.
We also require the energy-weighted EM cluster width in the finely-segmented third EM layer
to be consistent with that expected for an electromagnetic shower.
In addition to calorimeter isolation, we apply track isolation,
requiring that the scalar sum of the transverse momenta of tracks
in an annulus of $0.05 \leq {\cal R} \leq 0.4$, calculated
around the EM cluster direction, is less than 1.5~GeV.

Jets are reconstructed by clustering energy deposited
in the calorimeter towers using 
the iterative midpoint cone algorithm~\cite{Run2Cone}
with a cone size of $0.7$. Jets must satisfy
quality criteria that suppress background from leptons, photons, and
detector noise effects.
To reject background from cosmic rays and $W\to e \nu$ decays,
the missing transverse momentum, calculated  
as a vector sum of the transverse energies of all calorimeter cells, 
is required to be less than $0.7 \cdot p_{T}^{\gamma}$.
All pairs of objects ($i, j$) in the event (for example, photon and jet or jet and jet) are required to be
separated by $\Delta{R}=\sqrt{(\Delta\eta_{ij})^{2}+(\Delta\phi_{ij})^2}>0.9$.

Each event must contain at least one photon in the pseudorapidity
region $|\eta_{\rm det}|<1.0$ or $1.5<|\eta_{\rm det}|<2.5$ 
and at least two (or three) jets with $|\eta_{\rm det}|<3.5$.
Events are selected with photon transverse momentum 
$50<p^{\gamma}_{T}<90$~GeV, leading jet $p_T>30$~GeV, while
the next-to-leading (second) jet must have $p_T>15$~GeV.
If there is a third jet with $p_T>15$ GeV that passes the selection criteria, the event
is also considered for the $\gamma + 3$ jet analysis.
The higher $p^{\gamma}_{T}$ scale (i.e., the scale of the first parton interaction),
compared to the lower $p_T$ threshold required of the second (and eventual third) jet,
results in a good separation between the first and second parton interactions of a DP event in momentum space.
The reconstructed energy of each jet formed from calorimeter
energy depositions does not correspond to the actual energy
of the jet particles which enter the calorimeter.
It is therefore corrected for the energy response of the
calorimeter, energy showering in and out the jet cone,
and additional energy from event pile-up and multiple $p\bar{p}$ interactions.

The sample of DP candidates is selected from events with 
a single reconstructed $p\bar{p}$ collision vertex. 
The collision vertex is required to have at least three associated tracks 
and to be within 60~cm of the center of the detector in the coordinate along the beam ($z$) axis. 
The probability for any two $p\bar{p}$ collisions occurring in the same bunch crossing 
for which a single vertex is reconstructed was estimated in \cite{D0_2010} and found to be $<\!10^{-3}$.

\section{Single and multiple interaction models}
\label{Sec:Models}

Monte Carlo (MC) samples are used for two purposes in this analysis.
First, we use them to calculate reconstruction efficiencies
and to unfold the data spectra to the particle level.
Secondly, differential cross sections of \gpj events simulated using different MPI models as
implemented in the {\sc pythia} and {\sc sherpa} \cite{Sherpa} event generators
are compared with the measured cross sections. 

There are two main categories of MPI models based on different sets of data used in the determination
of the models parameters (it is customary to refer to different models
and to their settings as ``tunes'').
The two categories, ``old'' and ``new,'' correspond to different approaches in the treatment
of MPI, ISR and FSR, and other effects \cite{Perugia,pT_order}.
The main difference between the ``new'' \cite{PYT} and the ``old'' models is
the implementation of the interplay between MPI and ISR, 
i.e., considering these two effects in parallel, in a common sequence of decreasing
$p_T$ values.
In the ``old'' models, 
ISR and FSR were included only for the hardest
interaction, and this was done before any additional interactions were 
considered. In the ``new'' models, all parton interactions include ISR and FSR
separately for each interaction. 
The new models, especially those corresponding to the Perugia family of tunes \cite{Perugia}, also allow for
a much wider set of physics processes to occur in the additional interactions.
A detailed description of the different {\sc pythia} MPI models can be found elsewhere \cite{PYT,Perugia}.
Here we provide a brief description of the models considered in our analysis.
They include: Perugia-0 (P0, the default model in the Perugia family \cite{Perugia});
P-hard and P-soft, which explore the dependence on
the strength of ISR/FSR effects, while maintaining a roughly-consistent MPI 
model as implemented in the P0 tune; P-nocr, which excludes any color reconnections in the final state; 
P-X and P-6, which are P0 modifications based on the MRST LO* and CTEQ6L1 PDF sets, 
respectively
(P0 uses CTEQ5L as a default).
We also compare data with predictions determined using tunes A and DW 
as representative of the ``old'' MPI models.

The measured cross sections are also compared with predictions obtained from the {\sc sherpa}
event generator, 
which also contains a simulation of MPI. 
Its initial modelling was similar to tune A from {\sc pythia} \cite{TH3}, but 
it has evolved and now is characterized, in particular, by 
(a) showering effects 
in the second interaction
and (b) a combination of the CKKW merging approach with the MPI modeling \cite{Sherpa,CKKW}.
Another distinctive feature of {\sc sherpa} is the modeling of the parton-to-photon fragmentation
contributions through the incorporation of QED effects into the parton shower \cite{Sherpa_photons}.

The data are also compared with models without MPI, in which the photon and all the jets are
produced exclusively in SP scattering.
Such events are simulated in both {\sc pythia} and {\sc sherpa}.
In {\sc pythia}, only $2\to 2$ diagrams are simulated, resulting in the production of a photon and 
a leading jet. With the MPI event generation switched off, all the additional (to the leading) jets 
are produced in the parton shower development in the initial and final states.
We refer to such SP events as ``{\sc pythia} SP'' events.
In {\sc sherpa}, up to two extra partons (and thus jets) are allowed at the matrix element level
in the $2\to \{2,3,4\}$ scattering, but jets can also be produced in parton showers. 
To provide a matching between the matrix-element partons and parton shower jets, we follow the
recommendation provided in \cite{Sherpa_photons} and choose the following
``matching'' parameters:
the energy scale $Q_0=30$ GeV and the spatial scale $D=0.4$, where $D$ 
is taken to be of the size of the photon isolation cone~\cite{Sherpa_matching}. 
This is the default scheme for the production
of \gpj events with and without MPI simulation. 
The set of events produced without MPI simulation within this scheme is called ``{\sc sherpa-1} SP''.
We study the dependence of the measured DP fractions on the scale choice in the {\sc sherpa} SP models
in Sec. \ref{Sec:dpfrac} by setting the matching scale $Q_0$ equal to 20 and to 40 GeV (sets ``{\sc sherpa-2}'' and 
``{\sc sherpa-3}'' respectively).
For completeness we consider {\sc sherpa} SP events in which all of the extra jets are produced (as in {\sc pythia})
in the parton shower with only a $2\to 2$ matrix element, and call this set ``{\sc sherpa-4} SP''.

\section{Data analysis and corrections}
\label{Sec:Anal}

\subsection{Background studies}

The main background to isolated photons comes from jets in which
a large fraction of the transverse momentum is carried by photons
from $\pi^0$, $\eta$, or $K^0_s$ decays. 
The photon-enhancing criteria described in Sec. \ref{Sec:Sel} are developed 
to suppress this background. 
The normalized $\Delta S$ and $\Delta \phi$ distributions are not very sensitive to the exact amount
of background from events with misidentified photons.
To estimate the photon fractions in the \dPhi bins,
we use the output of two neural networks (NN) \cite{gamjet_xs}.
These NNs are constructed using the {\sc jetnet} package \cite{JN} and are trained 
to discriminate between photon and EM-jets in the central and end calorimeter regions
using calorimeter shower shape and track isolation variables \cite{gamjet_xs}. 
The distribution of the photon NN output for the simulated photon
signal and for the dijet background samples are fitted to
data in each \dPhi bin using a maximum likelihood fit~\cite{HMCMLL} 
to obtain the fractions of signal events in the data.
To obtain a more statistically significant estimate of the photon purity in the \dPhi bins,
we use a single $\ptsj$ bin: $15 < \ptsj < 30$ GeV.
The fit results show that the \gpj signal fractions in all  \dPhi bins agree well  with a constant value, 
$0.69\pm0.03$ in the central and $0.71\pm0.02$ in the end calorimeter regions. 

The sensitivity of the $\Delta S$ and $\Delta \phi$ distributions to this background from jets
is also examined by considering two data samples in addition to the sample with
the default photon selections:
one with relaxed and another with tighter track and calorimeter isolation requirements.
According to MC estimates, in those two samples the fraction of background events
should either increase or decrease by ($30-35$)\% with respect to the default sample.
We study the variation of the $\Delta S$ and $\Delta \phi$ normalized cross sections in data
by comparing the relaxed and tighter data sets 
with the default set. We find that the cross section variations are within 5\%.

\subsection{Efficiency and unfolding corrections}
\label{Sec:Unf}

\begin{figure}[htbp]
~\\[-0mm]
\hspace*{0mm} \includegraphics[scale=0.32]{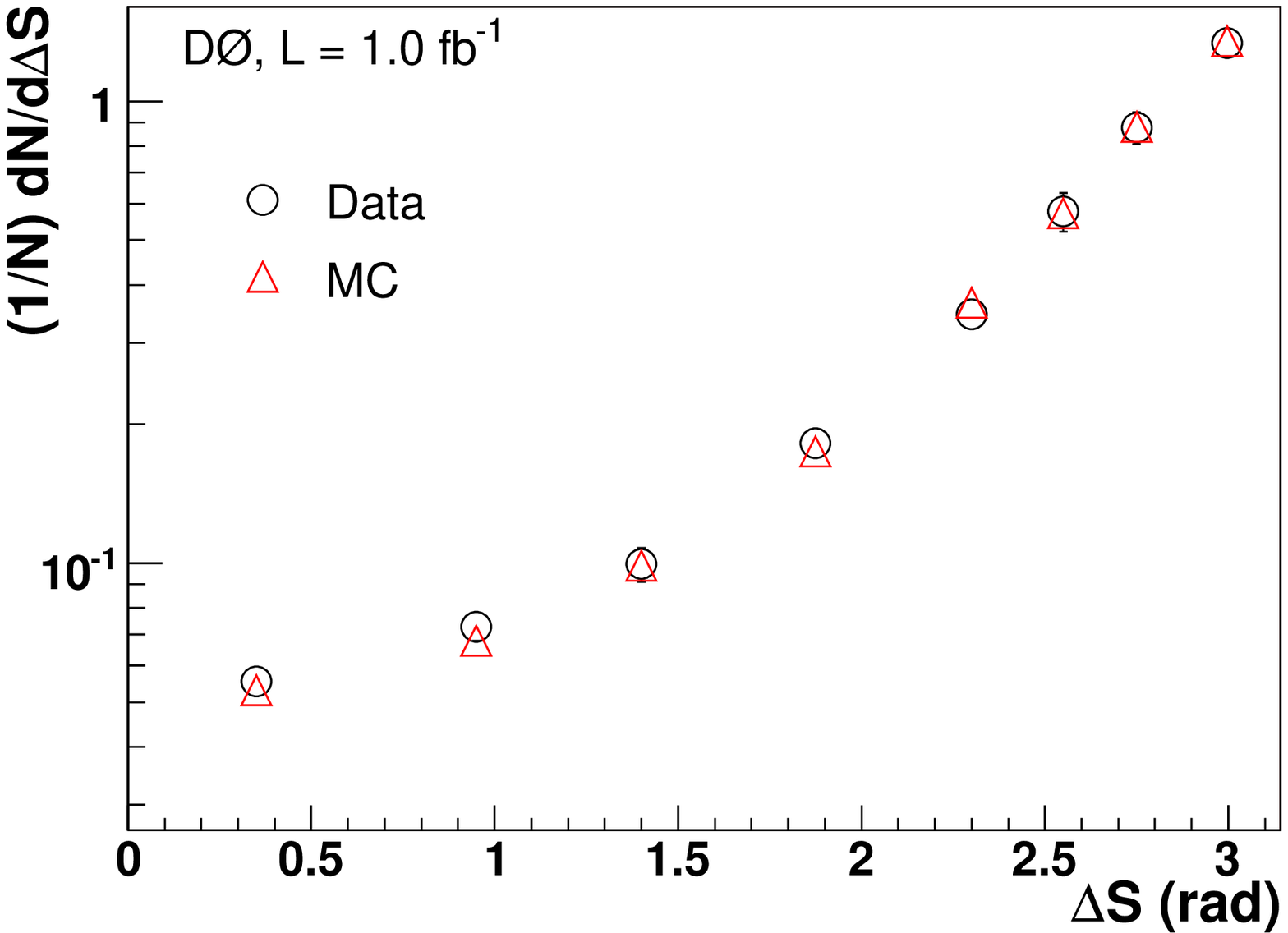}
~\\[-2mm]
\caption{Normalized $\Delta S$ distribution for data and for the reweighted MC sample in the range $15<\ptsj<30$ GeV.}
\label{fig:ds_reco}
~\\[-0mm]
\hspace*{0mm} \includegraphics[scale=0.32]{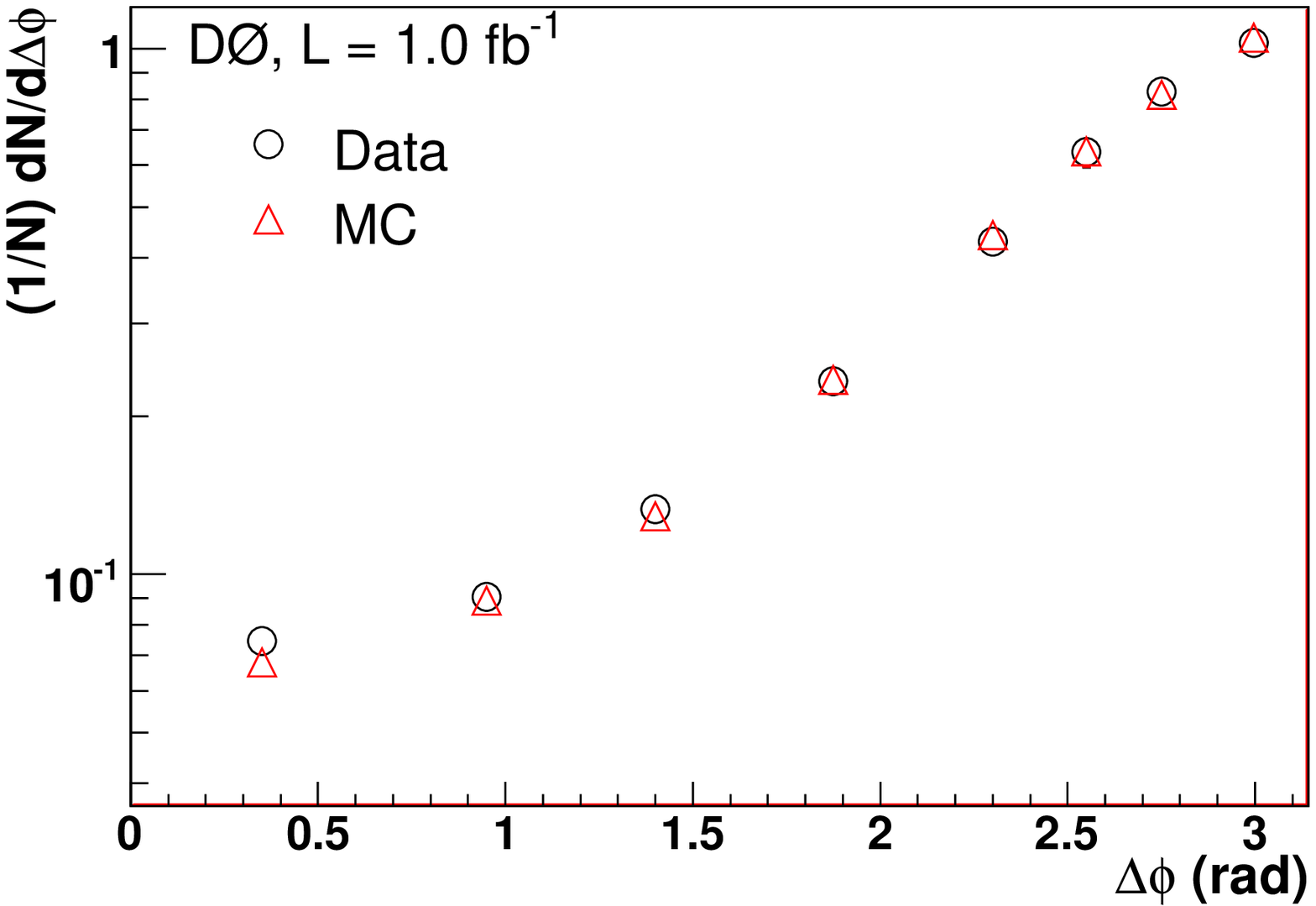}
~\\[-2mm]
\caption{Normalized $\Delta \phi$ distribution for data and for the reweighted MC sample in the range $15<\ptsj<20$ GeV.}
\label{fig:dphi1_reco}
\end{figure}

To select \gpTWOj and \gpTHRj events, we apply the selection criteria described in Sec.~\ref{Sec:Sel}.
The selected events are then corrected for selection efficiency, acceptance, and event
migration effects in bins of $\Delta S$ and $\Delta \phi$.
These corrections are calculated using MC events generated using {\sc pythia}
with tune P0, as discussed in Sec. \ref{Sec:Models}.
The  generated MC events are processed through a {\sc geant}-based~\cite{GEANT} simulation of the D0 detector
response. 
These MC events are then processed using the same reconstruction code as used for the data.
We also apply additional smearing to the reconstructed photon and jet $p_T$
so that the resolutions in MC match those observed in data.
The reconstructed $\Delta S$ and $\Delta \phi$ distributions 
in the simulated events using the P0 tune are found to describe the data.
In addition to the simulation with the default tune P0, 
we have also considered P0 MC events that have been reweighted 
to reproduce the $p_T$ distributions in data. 
After such reweighting, the reconstructed  $\Delta S$ and $\Delta \phi$ distributions 
give an excellent description of the data. Figures~\ref{fig:ds_reco} and \ref{fig:dphi1_reco}
show the normalized distributions as a function of $\Delta S$ for \gpTHRj and $\Delta \phi$ for \gpTWOj events 
(for the $\ptsj$ bin $15-20$ GeV, chosen as an example) 
in data and in the reweighted MC.

Three sets of corrections are applied in data to obtain the differential cross sections
which we then compare with the various MPI models.
We apply them to correct for detector and reconstruction inefficiencies
and for bin migration effects.
The first correction deals with the possibility 
that, due to the detector and reconstruction effects, our selected event sample may contain events 
which would fail the selection criteria at the particle level.
The data distributions are corrected, on a bin-by-bin basis, for the fraction of events of this type.
We also apply a correction for events which fail the selection requirement at the reconstruction level.
Systematic uncertainties are assigned on these two correction factors to account for uncertainties
on the photon and jet identification, the jet energy scale (JES) and resolution, and vary in the 
$\Delta S$ ($\Delta \phi$) bins up to 12\% (18\%) in total.
They are dominated by the JES uncertainties.
The overall corrections obtained with the default P0 and with the reweighted MC samples 
agree within about 5\% for most 
$\Delta S$ and $\Delta \phi$ bins and differ by at most 25\%.
Since we are measuring normalized
cross sections, the absolute values of the corrections 
are not important, and we need only their relative dependence 
on $\Delta S$ and $\Delta \phi$.

The third correction accounts for the migration of events between different bins
of the $\Delta S$ and $\Delta \phi$ distributions, which is caused by the finite photon and
jet angular resolutions and by energy resolution effects, and can change the $p_T$ ordering
of jets between the reconstruction and the particle level.
To obtain the $\Delta S$ and $\Delta \phi$ distributions at the particle level, we follow
the unfolding procedure described in the Appendix, based on the 
Tikhonov regularization method \cite{Tikhonov,Anikeev,Anikeev1,Cowan}.
The bin sizes for the $\Delta S$ and $\Delta \phi$ distributions are 
chosen to have sensitivity to different MPI models 
(which is largest for small $\Delta S$ and $\Delta \phi$ angles)
while keeping good statistics and the bin-to-bin migration small.
The statistical uncertainties ($\delta_{\rm stat}$) are
in the range $(10-18)\%$. They are due to the procedure of regularized unfolding 
and take into account the correlations between the bins.
The correlation factor for adjacent bins in the unfolded distributions is about $(30-45)\%$,
and it is reduced to $\approx 10\%$ for other (next-to-adjacent)
bins.
To validate the unfolding procedure, a MC closure test is
performed.  We compare the unfolded MC distribution to the true
MC distribution and find that they agree within statistical uncertainties.

\section{Differential cross sections and comparison with models}
\label{Sec:xs_data_theory} 

In this section, we present the four measurements of normalized differential cross sections, \DSigDS
in a single $\ptsj$ bin ($15-30$ GeV) for \gpTHRj events and
\DSigDPhi in three $\ptsj$ bins ($15-20$, $20-25$, and $25-30$ GeV) for \gpTWOj events.
The results are presented numerically in Tables \ref{tab:xs_ds} -- \ref{tab:xs_dphi3}
as a function of $\Delta S$ and $\Delta \phi$, with the bin centers 
estimated using the theoretical predictions obtained using the P0 tune.
The column ``$N_{\rm data}$'' shows the number of selected data events in each 
$\Delta S$ ($\Delta \phi$) bin at the reconstruction level.
The differential distributions decrease by about two orders of magnitude when moving from
the $\Delta S$ ($\Delta \phi$) bin $2.85-3.14$ radians to the bin $0.0-0.7$ radians 
and have a total uncertainty ($\delta_{\rm tot}$)
between 7\% and 30\%. Here $\delta_{\rm tot}$ is defined as a sum in quadrature of statistical
($\delta_{\rm stat}$) and systematical ($\delta_{\rm syst}$) uncertainties.
It is dominated by systematic uncertainties.
The sources of systematic uncertainties are 
the JES  $(2-17\%)$, largest at the small angles, unfolding ($5-18\%$),
jet energy resolution simulation in MC events ($1-7\%$), and background contribution (up to $5\%$).

\begin{figure}[htbp]
\hspace*{0cm} \includegraphics[scale=0.4]{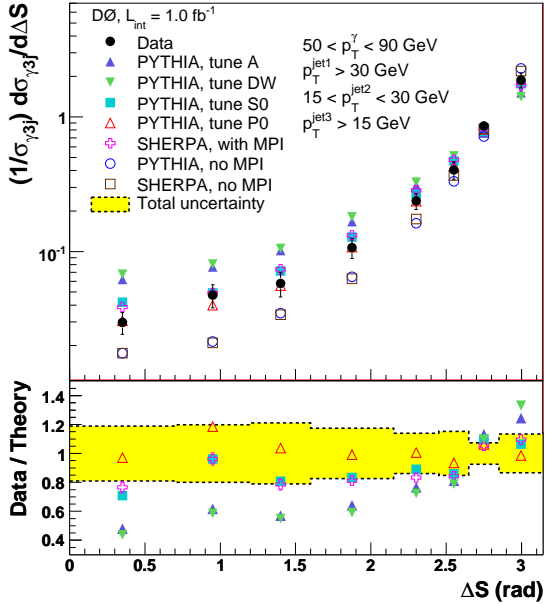}
\caption{Normalized differential cross section in \gpTHRj events, \DSigDS, in data compared to MC models and the ratio of data 
over theory, only for models including MPI, in the range $15 < \ptsj < 30$ GeV.}
\label{fig:ds_dt_th}
\end{figure}

\begin{figure}[htbp]
\includegraphics[scale=0.4]{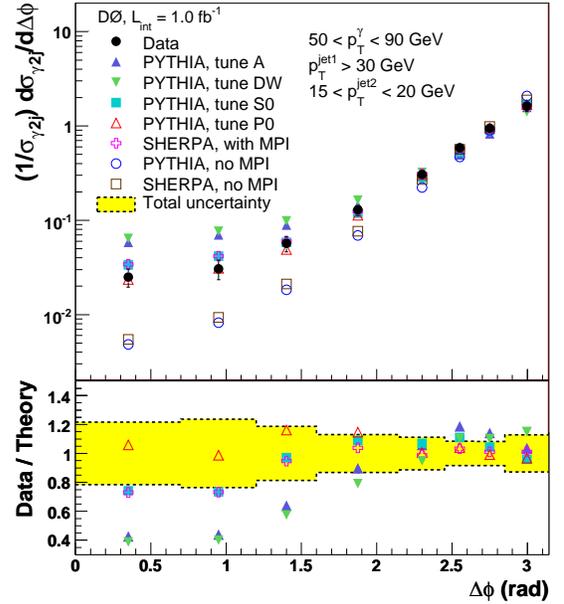}
\caption{Normalized differential cross section in \gpTWOj events, \DSigDPhi, in data compared to MC models and the ratio of data 
over theory, only for models including MPI, in the range $15 < \ptsj < 20$ GeV.}
\label{fig:dphi1_dt_th}
\end{figure}

\begin{figure}[htbp]
\includegraphics[scale=0.4]{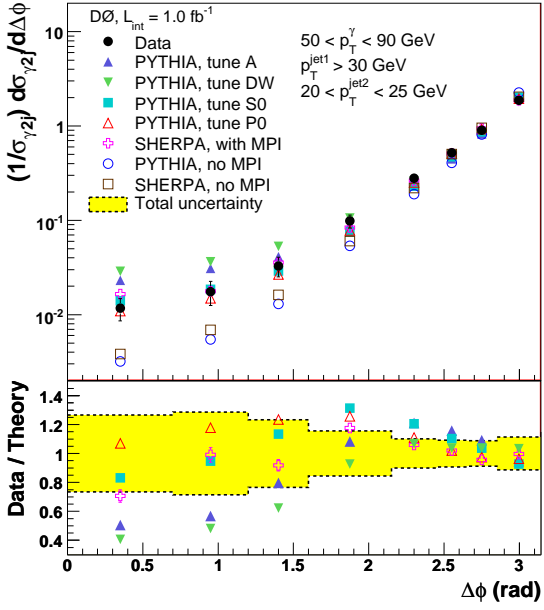}
\caption{Normalized differential cross section in \gpTWOj events, \DSigDPhi, in data compared to MC models and the ratio of data 
over theory, only for models including MPI, in the range $20 < \ptsj < 25$ GeV}.
\label{fig:dphi2_dt_th}
\end{figure}

\begin{figure}[htbp]
\includegraphics[scale=0.4]{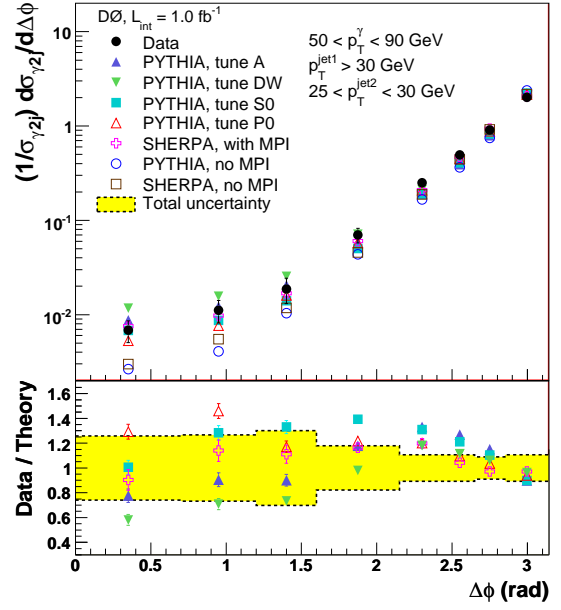}
\caption{Normalized differential cross section in \gpTWOj events, \DSigDPhi, in data compared to MC models and the ratio of data 
over theory, only for models including MPI, in the range $25 < \ptsj < 30$ GeV.}
\label{fig:dphi3_dt_th}
\end{figure}

The results are compared in Figs.~\ref{fig:ds_dt_th}--\ref{fig:dphi3_dt_th} to predictions from different
MPI models implemented in {\sc pythia} and {\sc sherpa}, as discussed in Sec.~\ref{Sec:Models}.
We also show predictions of SP models 
in {\sc pythia} and {\sc sherpa} ({\sc sherpa-1} model). 
In the QCD NLO predictions, only final states with a \gpTWOj topology are considered, 
and thus for the direct photon production diagrams, we should have $\DPhi = \pi$. 
The $\DPhi$ angle may differ from $\pi$ due to photon production
through a parton-to-photon fragmentation mechanism. 
Even if we take into account this production mechanism, which is included in the {\sc jetphox} \cite{JETPHOX} 
NLO QCD calculations, only the two highest $\Delta\phi$ bins receive significant contributions.

Figs.~\ref{fig:ds_dt_th}--\ref{fig:dphi3_dt_th} show the sensitivity of the two angular variables 
$\Delta S$ and $\Delta\phi$ to the various MPI models,
with predictions varying significantly and differing from each other by
up to a factor 2.5 at small $\Delta S$ and $\Delta\phi$,
in the region where the relative DP contribution is expected to be the highest.
The sensitivity is reduced by the choice of SP model,
for which we derive an upper value of 25\% comparing the ratios of predictions from various
models ({\sc pythia}, {\sc sherpa-2, -3, -4}). This upper value is considerably smaller
than the difference between the various MPI models.

Tables \ref{tab:chi2_1} and \ref{tab:chi2_2} are complementary to 
Figs.~\ref{fig:ds_dt_th} -- \ref{fig:dphi3_dt_th} and show the $\chi^2/ndf$ values
of the agreement between theory and data for each model. 
Here $ndf$ stands for the number of degrees of freedom 
(taken as the number of bins, $N_{\rm bins}$ minus 1),
and $\chi^2$ is calculated as 
\vskip-4mm
\begin{equation}
\chi^2=\sum_{i=1}^{N_{\rm bins}}\frac{(D_i-T_i)^2}{\delta^2_{i, {\rm unc}}}, 
\label{eq:chi2}
\end{equation}
where $D_i$ and $T_i$ represent the cross section values in the $i$-th bin of data and a theoretical
model respectively, while $\delta^2_{i,{\rm unc}}$ is the total uncorrelated uncertainty in this bin. 
The latter is composed of the uncertainties for the corrections in the unfolding procedure 
(Sec.~\ref{Sec:Unf}), the statistical uncertainties of the data $\delta_{\rm stat}$ and the 
theoretical model. 
The uncorrelated uncertainty $\delta^2_{i, {\rm unc}}$ is always larger than
all remaining correlated systematic uncertainties.
Since small angles ($\Delta S (\Delta\phi) \lesssim 2$) are the most sensitive to DP contributions,
we calculate the $\chi^2/ndf$ separately for these bins.
From Figs.~\ref{fig:ds_dt_th} -- \ref{fig:dphi3_dt_th} and Tables \ref{tab:chi2_1} and \ref{tab:chi2_2},
we conclude:
(a) the predictions derived from SP models do not describe the measurements;
(b) the data favor the predictions of the new {\sc pythia} MPI models (P0, P-hard, P-6, P-X, P-nocr)
and to a lesser extent S0 and {\sc sherpa} with MPI; 
and (c) the predictions from tune A and DW MPI models are disfavored.
\begin{table}[h]
\begin{center}
\caption{Measured normalized differential cross sections \DSigDS for $15 < \ptsj < 30$ GeV.}
\label{tab:xs_ds}
\begin{tabular}{ccrcrrr} \hline\hline
~~$\Delta S$ bin~  & $~~~\la\Delta S\ra$ & $N_{\rm data}$ & Normalized & \multicolumn{3}{c}{Uncertainties (\%)} \\
    (rad)          & ~~~(rad)            &  & cross section & $\delta_{\rm stat}$ & $\delta_{\rm syst}$ & $\delta_{\rm tot}$ \\\hline
0.00 -- 0.70 &  ~~~0.36 & 495 & $2.97 \times 10^{-2}$ &  11.3 &  14.7 &  18.6 \\
0.70 -- 1.20 &  ~~~0.97 & 505 & $4.74 \times 10^{-2}$ &  12.3 &  15.6 &  19.9 \\
1.20 -- 1.60 &  ~~~1.42 & 498 & $5.80 \times 10^{-2}$ &  13.4 &  15.8 &  20.7 \\
1.60 -- 2.15 &  ~~~1.90 & 1315 & $1.11 \times 10^{-1}$ &   7.5 &  15.3 &  17.0 \\
2.15 -- 2.45 &  ~~~2.32 & 1651 & $2.38 \times 10^{-1}$ &   6.0 &  12.0 &  13.4 \\
2.45 -- 2.65 &  ~~~2.56 & 1890 & $4.04 \times 10^{-1}$ &   5.6 &  13.6 &  14.7 \\
2.65 -- 2.85 &  ~~~2.76 & 3995 & $8.59 \times 10^{-1}$ &   3.2 &   5.6 &   6.4 \\
2.85 -- 3.14 &  ~~~3.02 &12431 & $\!\!\!\!1.89 \times 10^{0}$ &   1.0 &  13.0 &  13.0 \\\hline\hline
\end{tabular}
\end{center}
\end{table}
\begin{table}[h]
\begin{center}
\caption{Measured normalized differential cross sections \DSigDPhi for $15 < \ptsj < 20$ GeV.}
\label{tab:xs_dphi1}
\begin{tabular}{ccrcrrr} \hline\hline
~~$\Delta \phi$ bin~  & ~~~$\la\Delta \phi\ra$ & $N_{\rm data}$ & Normalized & \multicolumn{3}{c}{Uncertainties (\%)} \\
    (rad)          & ~~~(rad)            & & cross section & $\delta_{\rm stat}$ & $\delta_{\rm syst}$ & $\delta_{\rm tot}$ \\\hline
0.00 -- 0.70    &  ~~~0.36 & 1028 & $2.49 \times 10^{-2}$ &   9.4 &  19.1 &  21.3 \\
0.70 -- 1.20    &  ~~~0.96 & 822 &  $3.06 \times 10^{-2}$&  11.8 &  20.3 &  23.4 \\
1.20 -- 1.60    &  ~~~1.42 & 1149 & $5.68 \times 10^{-2}$ &   9.6 &  15.5 &  18.2 \\
1.60 -- 2.15 	&  ~~~1.92 & 3402 & $1.29 \times 10^{-1}$ &   4.9 &  11.5 &  12.5 \\
2.15 -- 2.45 	&  ~~~2.32 & 4187 & $3.06 \times 10^{-1}$ &   4.5 &   9.5 &  10.5 \\
2.45 -- 2.65 	&  ~~~2.56 & 5239 & $5.88 \times 10^{-1}$ &   4.0 &   6.3 &   7.4 \\
2.65 -- 2.85 	&  ~~~2.76 & 8246 & $9.43 \times 10^{-1}$ &   3.0 &   6.8 &   7.5 \\
2.85 -- 3.14 	&  ~~~3.01 & 20337 & $\!\!\!\!1.63 \times 10^{0}$ &   1.1 &  12.3 &  12.3 \\\hline\hline
\end{tabular}
\end{center}
\end{table}
\begin{table}[h]
\begin{center}
\caption{Measured normalized differential cross section \DSigDPhi for $20 < \ptsj < 25$ GeV.}
\label{tab:xs_dphi2}
\begin{tabular}{ccrcrrr} \hline\hline
~~$\Delta \phi$ bin~  & ~~~$\la\Delta \phi\ra$ & $N_{\rm data}$ & Normalized & \multicolumn{3}{c}{Uncertainties (\%)} \\
    (rad)          & ~~~(rad)            & & cross section & $\delta_{\rm stat}$ & $\delta_{\rm syst}$ & $\delta_{\rm tot}$ \\\hline
0.00 -- 0.70 &  ~~~0.35 & 388 & $1.17 \times 10^{-2}$ &  12.5 &  23.2 &  26.4 \\
0.70 -- 1.20 &  ~~~0.96 & 358 & $1.75 \times 10^{-2}$ &  17.7 &  22.2 &  28.5 \\
1.20 -- 1.60 &  ~~~1.42 & 489 & $3.29 \times 10^{-2}$ &  15.6 &  17.0 &  23.1 \\
1.60 -- 2.15 &  ~~~1.92 & 1848 & $9.84 \times 10^{-2}$ &   6.2 &  13.8 &  15.1 \\
2.15 -- 2.45 &  ~~~2.33 & 2682 & $2.80 \times 10^{-1}$ &   4.6 &   8.2 &   9.4 \\
2.45 -- 2.65 &  ~~~2.56 & 3208 & $5.21 \times 10^{-1}$ &   4.5 &   7.1 &   8.4 \\
2.65 -- 2.85 &  ~~~2.77 & 5404 & $9.01 \times 10^{-1}$ &   3.2 &   7.3 &   8.0 \\
2.85 -- 3.14 &  ~~~3.02 & 15901 & $\!\!\!\!1.88 \times 10^{0}$ &   1.0 &  10.8 &  10.8 \\\hline\hline
\end{tabular}
\end{center}
\end{table}
\begin{table}[h]
\begin{center}
\caption{Measured normalized differential cross section \DSigDPhi for $25 < \ptsj < 30$ GeV.}
\label{tab:xs_dphi3}
\begin{tabular}{ccrcrrr} \hline\hline
~~$\Delta \phi$ bin~  & ~~~$\la\Delta \phi\ra$ & $N_{\rm data}$ & Normalized & \multicolumn{3}{c}{Uncertainties (\%)} \\
    (rad)          & ~~~(rad)            & & cross section & $\delta_{\rm stat}$ & $\delta_{\rm syst}$ & $\delta_{\rm tot}$ \\\hline
0.00 -- 0.70 &  ~~~0.32 & 158 & $6.82 \times 10^{-3}$ &  16.1 &  19.8 &  25.5 \\
0.70 -- 1.20 &  ~~~0.94 & 155 & $1.11 \times 10^{-2}$ &  20.9 &  16.4 &  26.6 \\
1.20 -- 1.60 &  ~~~1.45 & 190 & $1.87 \times 10^{-2}$ &  24.0 &  17.9 &  30.0 \\
1.60 -- 2.15 &  ~~~1.92 & 910 & $7.00 \times 10^{-2}$ &   7.0 &  15.9 &  17.4 \\
2.15 -- 2.45 &  ~~~2.32 & 1683 &  $2.50 \times 10^{-1}$ &   5.0 &   8.6 &   9.9 \\
2.45 -- 2.65 &  ~~~2.57 & 2155 &  $4.93 \times 10^{-1}$ &   4.5 &   8.9 &  10.0 \\
2.65 -- 2.85 &  ~~~2.77 & 3894 &  $9.09 \times 10^{-1}$ &   3.1 &   7.5 &   8.1 \\
2.85 -- 3.14 &  ~~~3.03 & 12332 &  $\!\!\!\!2.01 \times 10^{0}$ &   1.0 &  10.2 &  10.2 \\\hline\hline
\end{tabular}
\end{center}
\end{table}


\begin{table*}[]
\caption{The results of a $\chi^2$ test of the agreement between data points and theory predictions
for the $\Delta S$ (\gpTHRj) and $\Delta \phi$ (\gpTWOj)  distributions for 
$0.0 \leq \Delta S (\Delta \phi) \leq \pi$ rad. Values are $\chi^2/ndf$.}
\label{tab:chi2_1}
\begin{tabular}{ccrrrrrrrrrrrr} \hline\hline
 Variable & $\ptsj$ & \multicolumn{2}{c}{SP model} & \multicolumn{10}{c}{MPI model}  \\
  & (GeV)   & ~~{\sc pythia} &  {\sc sherpa} &   A~ &  ~DW &   ~~S0 &   ~~P0 &  P-nocr &  P-soft &  P-hard &  ~P-6 &   ~P-X &  {\sc sherpa} \\\hline
$\Delta S$    &$15 - 30$  & 7.7~~ & 6.0~~~   & 15.6 & 21.4 & 2.2 & 0.4 & 0.5~~ & 2.9~~ & 0.5~~ & 0.4 & 0.5 & 1.9~~~ \\
$\Delta \phi$ &$15 - 20$  & 16.6~~ & 11.7~~~ & 19.6 & 27.7 & 1.6 & 0.5 & 0.9~~ & 1.6~~ & 0.9~~ & 0.6 & 0.8 & 1.2~~~ \\
$\Delta \phi$ &$20 - 25$  & 10.2~~ & 5.9~~~  & 4.0 & 7.9 & 1.1 & 0.9 & 1.4~~   & 2.1~~ & 1.1~~ & 1.3 & 1.5 & 0.4~~~ \\
$\Delta \phi$ &$25 - 30$  & 7.2~~ & 3.5~~~   & 2.8 & 3.0 & 2.4 & 1.1 & 1.1~~   & 3.7~~ & 0.2~~ & 1.3 & 1.9 & 0.7~~~ \\\hline\hline
\end{tabular}
\end{table*}

\begin{table*}[]
\caption{The results of a $\chi^2$ test of the agreement between data points and theory predictions
for the $\Delta S$ (\gpTHRj) and the $\Delta \phi$ (\gpTWOj) distributions for $\Delta S (\Delta \phi) \leq 2.15$ rad. Values are $\chi^2/ndf$.}
\label{tab:chi2_2}
\begin{tabular}{ccrrrrrrrrrrrr} \hline\hline
 Variable & $\ptsj$ & \multicolumn{2}{c}{SP model} & \multicolumn{10}{c}{MPI model}  \\
  & (GeV)   & ~~{\sc pythia} &  {\sc sherpa} &   A~ &  ~DW &   ~~S0 &   ~~P0 &  P-nocr &  P-soft &  P-hard &  ~P-6 &   ~P-X &  {\sc sherpa} \\\hline
$\Delta S$    &$15 - 30$  & 10.9~~ & 11.3~~~ & 31.0 & 42.9 & 3.4 & 0.4 & 0.5~~ & 4.9~~ & 0.9~~ & 0.5 & 0.4 & 2.6~~~ \\
$\Delta \phi$ &$15 - 20$  & 30.2~~ & 26.0~~~ & 40.7 & 61.1 & 2.2 & 0.9 & 1.6~~ & 1.5~~ & 1.2~~ & 1.2 & 1.0 & 2.4~~~ \\
$\Delta \phi$ &$20 - 25$  & 15.4~~ & 12.1~~~ & 6.8 & 18.0 & 1.0 & 1.8 & 2.7~~ & 1.7~~ & 1.5~~ & 2.5 & 2.4 & 0.6~~~ \\
$\Delta \phi$ &$25 - 30$  & 7.1~~ & 5.3~~~   & 1.3 & 5.6 & 1.6 & 1.1 & 1.0~~ & 2.1~~ & 0.3~~ & 1.4 & 1.6 & 0.5~~~ \\\hline\hline
\end{tabular}
\end{table*}

\section{Fractions of double parton events in the $\gamma+2$ jet final state}
\label{Sec:dpfrac}

The comparison of the measured cross section with models 
(Sec.~\ref{Sec:xs_data_theory}) shows clear evidence for DP scattering. 
We use the measurement of the differential cross section with respect to $\Delta \phi$  
and predictions for the SP contributions to this cross sections in different models to determine
the fraction of \gpTWOj events which originate from DP interactions as a function of the
second parton interaction scale ($\ptsj$) and of $\Delta \phi$. 
Due to ISR and FSR effects the $p_T$ balance vectors of each system may be non-zero
and have an arbitrary orientation with respect to each other~\cite{Fig9},
which leads to a uniform \dPhi distribution for DP events.

\begin{figure}[htbp]
\hspace*{-0.2cm} \includegraphics[scale=0.4]{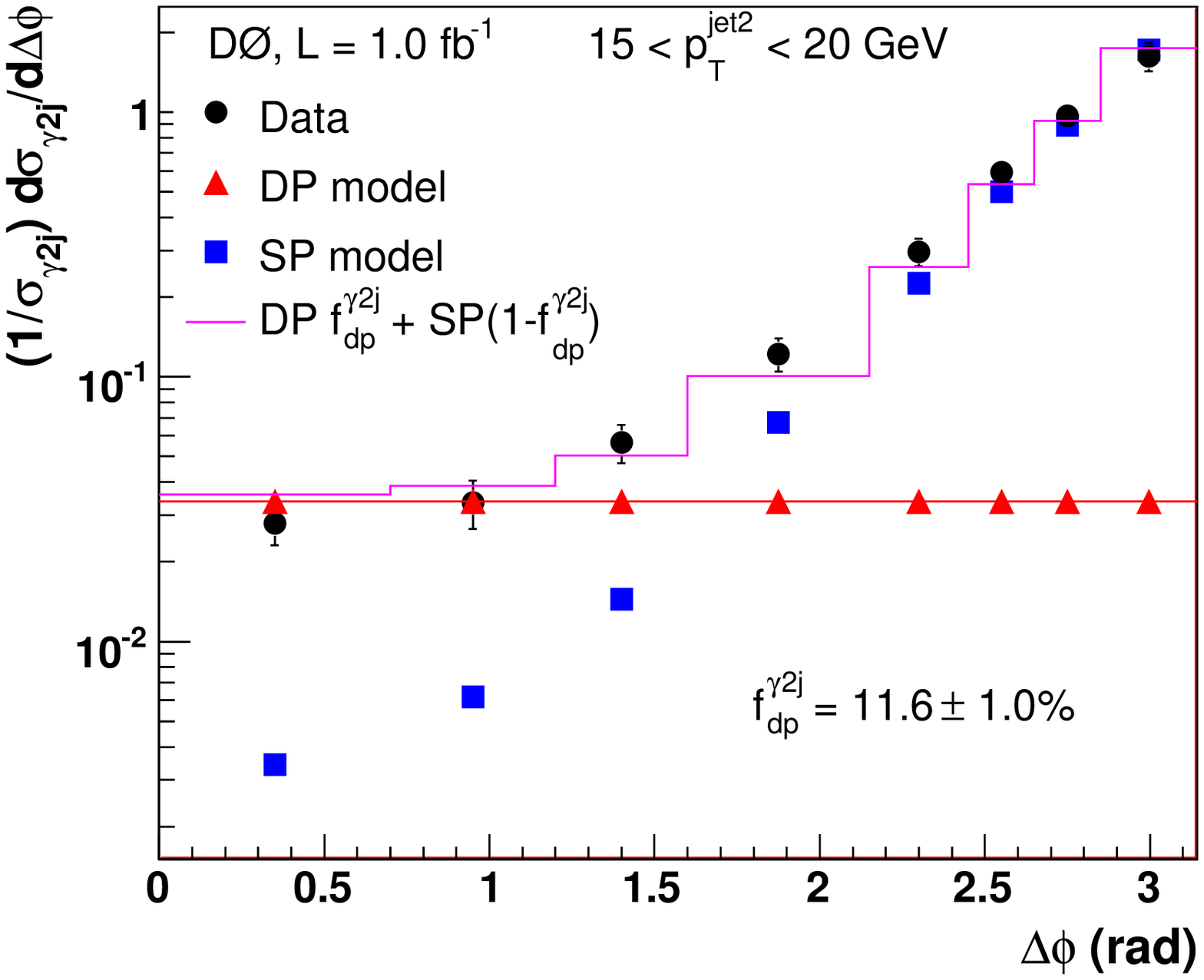}
\vskip -5mm
\caption{$\Delta \phi$ distribution in data, SP, and DP models, and the sum of the SP and DP contributions 
weighted with their fractions for $15<\ptsj<20$ GeV.}
\label{fig:dpfrac_dphi1}
\end{figure}

\begin{figure}[htbp]
\hspace*{-0.2cm} \includegraphics[scale=0.4]{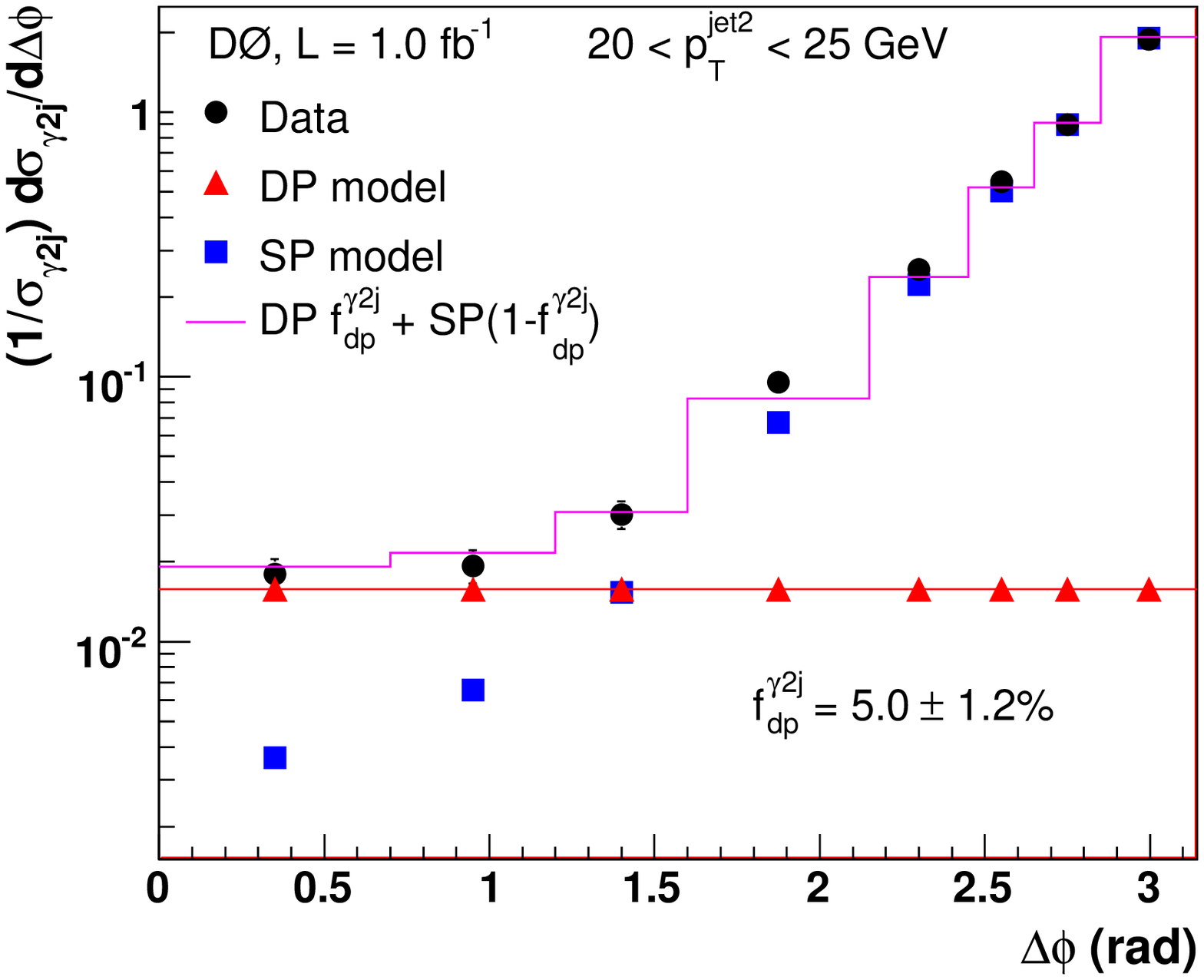}
\vskip -5mm
\caption{Same as in Fig.~\ref{fig:dpfrac_dphi1}, but for $20<\ptsj<25$ GeV.}
\label{fig:dpfrac_dphi2}
\end{figure}

\begin{figure}[htbp]
\hspace*{-0.2cm} \includegraphics[scale=0.4]{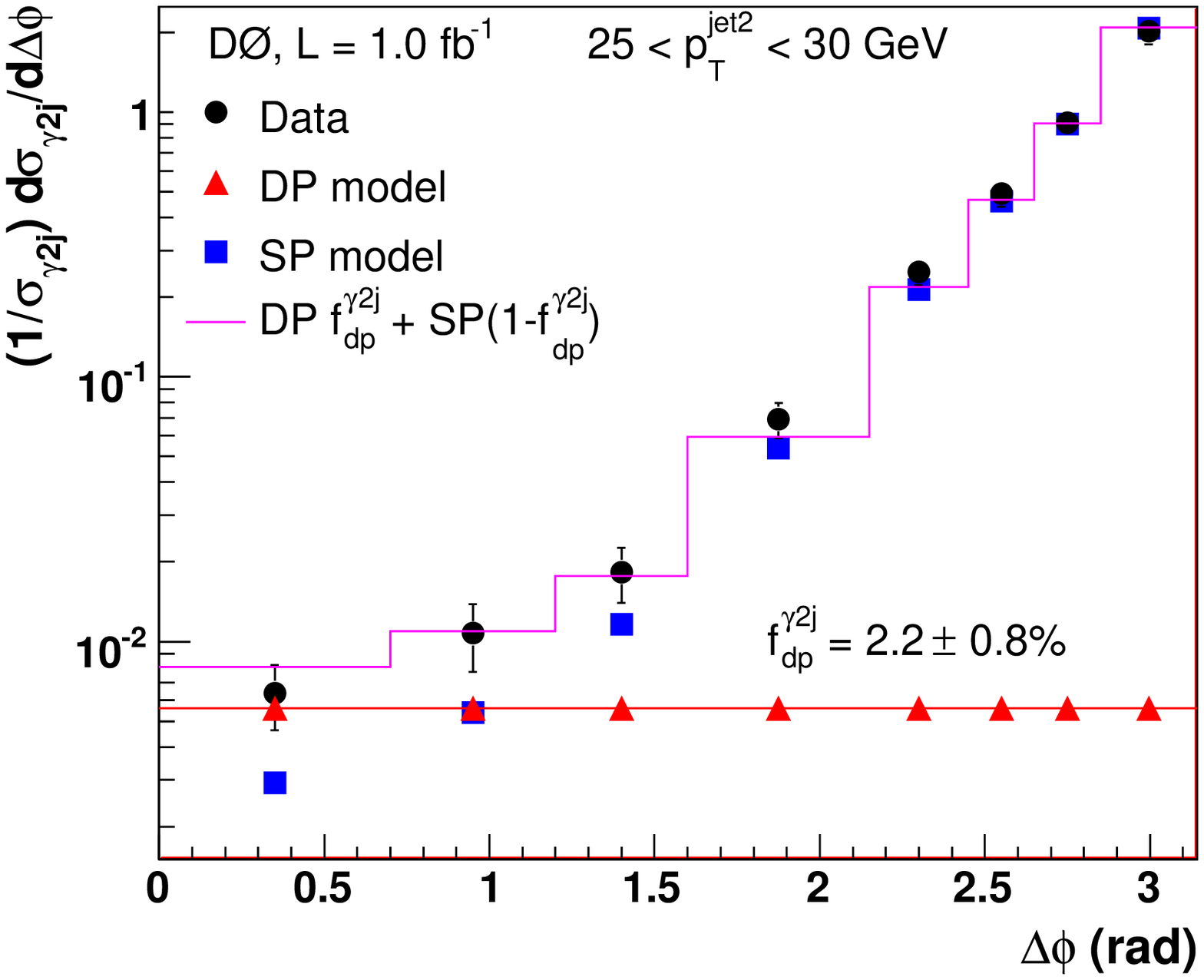}
\vskip -5mm
\caption{Same as in Fig.~\ref{fig:dpfrac_dphi1}, but for $25<\ptsj<30$ GeV.}
\label{fig:dpfrac_dphi3}
\end{figure}

Using the uniform distribution as the DP model template and 
the {\sc sherpa-1} prediction as the SP model template, we can fit the \dPhi distributions measured 
in data and obtain the fraction of DP events from a maximum likelihood fit \cite{HMCMLL}.
We repeat this procedure in three independent ranges of $\ptsj$.
The distributions in data, SP, and DP models, as well as a sum of the SP and DP distributions, 
weighted with their respective fractions, are shown in Figs.~\ref{fig:dpfrac_dphi1} -- \ref{fig:dpfrac_dphi3} 
for the  three $\ptsj$ intervals.
The sum of the SP and DP predictions reproduces the data well. 
The measured DP fractions ($f_{\rm dp}^{\gamma 2j}$) are presented in Table \ref{tab:dp_frac_tot}. 

The uncertainties in the DP fractions are due to the fit,
the total (statistical plus systematic) uncertainties on the data points, 
and the choice of SP model. The effect from the second source is estimated by varying
all the data points simultaneously up and down by the total experimental uncertainty ($\delta_{\rm tot}$).
The uncertainties due to the SP model are estimated by considering SP models (Sec.~\ref{Sec:Models})
that are different from the default
choice of {\sc sherpa-1}: 
{\sc -2}, {\sc -3}, {\sc -4}, as well as {\sc pythia} SP predictions. 
The measured DP fractions with all sources of uncertainties in each $\ptsj$ bin are summarized in Table \ref{tab:dp_frac_tot}.
The DP fractions in \gpTWOj events, $f_{\rm dp}^{\gamma 2j}$,  decrease as a function of $\ptsj$ from 
$(11.6\pm1.0)\%$ in the bin $15-20$ GeV, to $(5.0\pm1.2)\%$ in the bin $20-25$ GeV, and $(2.2\pm0.8)\%$ in the 
bin $25-30$ GeV.
The estimated DP fraction in \gpTWOj events 
selected with $\Ptg>16$ GeV and $p_T^{\rm jet}>8$ GeV from the CDF Collaboration \cite{CDF_gj_angles}
is $14^{+8}_{-7}\%$, 
which is in qualitative agreement with an extrapolation of our measured DP fractions to lower jet $p_T$.

\begin{figure}[htbp]
\hspace*{0.0cm} \includegraphics[scale=0.44]{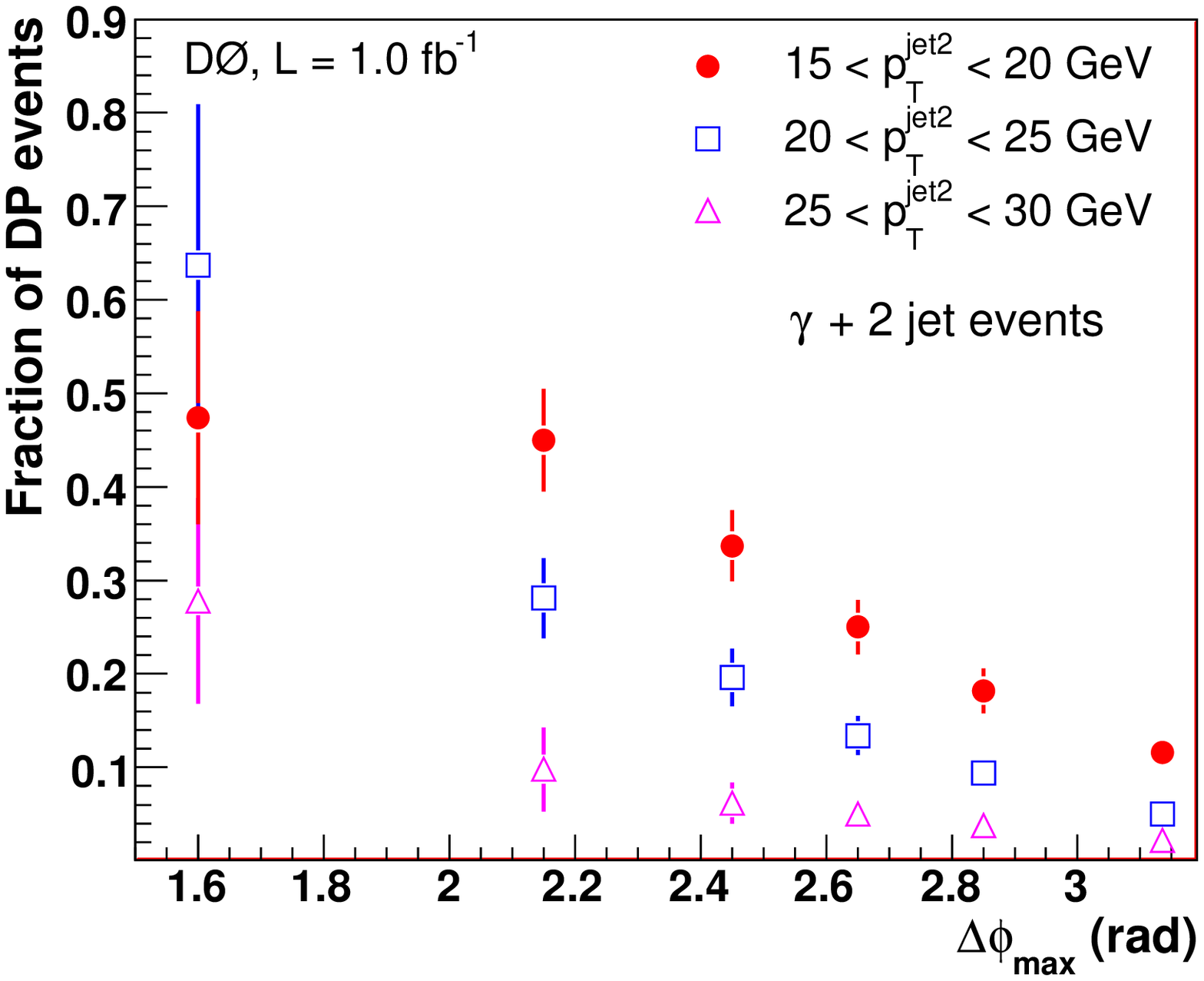}
\vskip -3mm
\caption{Fractions of DP events with total uncertainties in \gpTWOj final state as a function of 
the upper limit on $\DPhi$ for the three $\ptsj$ intervals.}
\label{fig:dpfrac_phimax}
\end{figure}

The DP fractions shown in Table \ref{tab:dp_frac_tot} are integrated over the entire region $0\leq\Delta\phi\leq\pi$.
However, from Figs.~\ref{fig:dpfrac_dphi1} -- \ref{fig:dpfrac_dphi3}, the fraction of DP events is
expected to be higher at smaller \dPhi. To determine the fractions as a function of \dPhi, 
we perform a fit in the different \dPhi regions by excluding the bins at high \dPhi; specifically,
by considering the \dPhi regions $0-2.85$, $0-2.65$, $0-2.45$, $0-2.15$, and $0-1.60$.
The DP fractions for these \dPhi regions
are shown in Table \ref{tab:dpfrac_dphi} for the three $\ptsj$ intervals.
The DP fractions with total uncertainties as functions of the upper limit on \dPhi ($\Delta \phi_{\rm max}$) for all 
the $\ptsj$ bins are also shown in Fig.~\ref{fig:dpfrac_phimax}.
As expected, they grow significantly towards the smaller angles and are higher for smaller $\ptsj$ bins.


\begin{table}[htpb]
\small
\caption{Fractions of DP events (\%)  with total uncertainties for $0\leq\DPhi\leq\pi$ in the three $\ptsj$ bins.}
\label{tab:dp_frac_tot}
\begin{tabular}{cccrrc} \hline\hline
$\ptsj$  ~&~ $\la \ptsj \ra$ ~&~ $f_{\rm dp}^{\gamma 2j}$ & \multicolumn{3}{c}{Uncertainties (in \%)} \\
 (GeV)   ~&~ (GeV) ~&~ (\%)          & ~~~~Fit   & $\delta_{\rm tot}$ & SP model\\\hline
$15-20$ ~&~ $17.6$ ~& $11.6\pm 1.4$ & 5.2   &  8.3   & ~6.7  \\
$20-25$ ~&~ $22.3$ ~&~ $5.0 \pm 1.2$ & 4.0   &  20.3  & 11.0  \\
$25-30$ ~&~ $27.3$ ~&~ $2.2 \pm 0.8$ & 27.8  &  21.0  & 17.9  \\\hline\hline 
\end{tabular}
\end{table}

\begin{table*}[]
\caption{DP fractions (\%) in data as a function of the $\DPhi$ interval for three $\ptsj$ bins.}
\label{tab:dpfrac_dphi}
\begin{tabular}{ccccccc} \hline\hline
 $\ptsj$  &  \multicolumn{6}{c}{$\DPhi$ interval (rad)}  \\
  (GeV)      & ~~~$0-\pi$~~ & ~~$0-2.85$~~ & ~~$0-2.65$~~ & ~~$0-2.45$~~ & ~~$0-2.15$~~ & ~~$0-1.6$~~ \\\hline
$15-20$ & $11.6\pm1.4$ & $18.2\pm 2.4$ & $25.0\pm 2.9$  & $33.7\pm3.8$ & $45.0\pm5.5$ & $47.4\pm11.4$ \\
$20-25$ & ~$5.0\pm1.2$ & ~$9.4\pm 1.2$ & $13.4\pm 2.1$  & $19.6\pm3.1$ & $28.1\pm4.3$ & $63.7\pm17.2$ \\
$25-30$ & ~$2.2\pm0.8$ & ~$3.8\pm 1.3$ & ~$5.0\pm 1.5$  & ~$6.2\pm2.2$ & ~$9.8\pm4.5$ & $27.8\pm11.5$  \\\hline\hline
\end{tabular}
\end{table*}

\section{Fractions of triple parton events in the $\gamma+3$~jet final state}
\label{Sec:tpfrac}

In this section, we estimate the fraction of \gpTHRj events from
triple parton interactions (TP) in data as a function of $\ptsj$.
In \gpTHRj TP events, the three jets come from three different
parton interactions, one \gpj and two dijet final states.
In each of the two dijet events, one of the jets is either not reconstructed or
below the 15 GeV $p_T$ selection threshold.

\begin{figure}[t]
\hspace*{0.1cm} \includegraphics[scale=0.17]{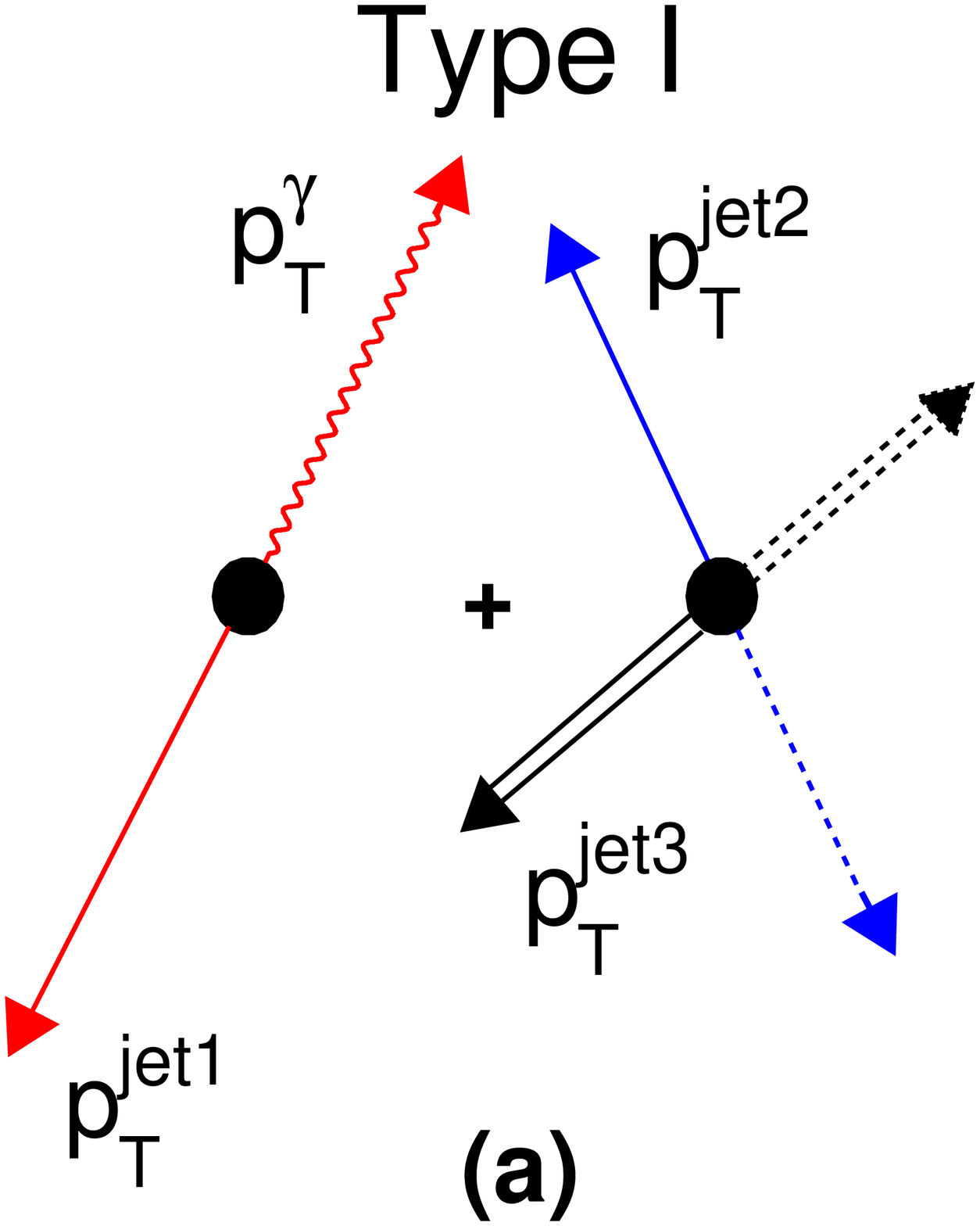}
\hspace*{0.2cm} \includegraphics[scale=0.17]{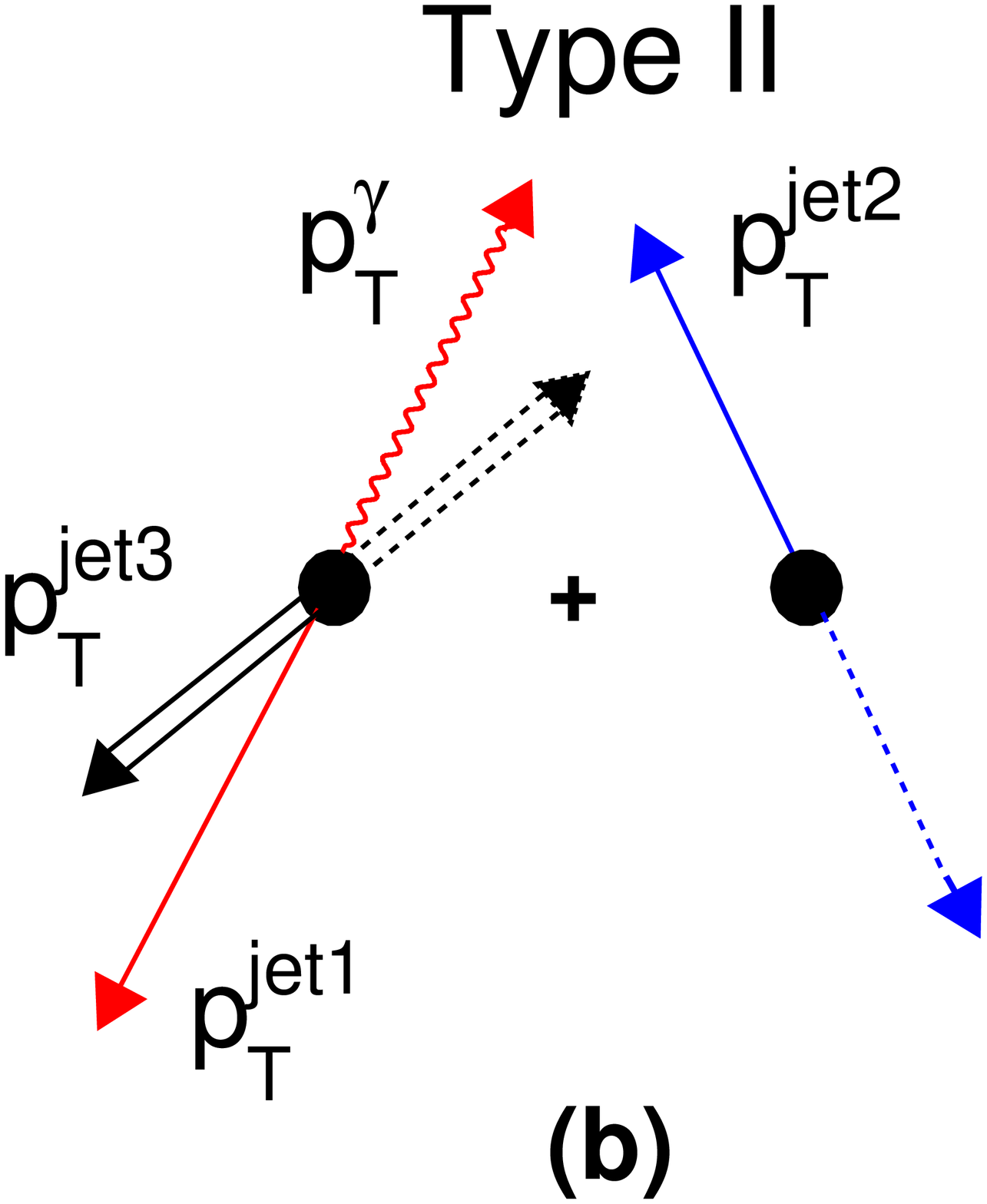}
\vspace*{-3mm} 
\caption{Two possible combinations of events present in the MixDP sample that actually represent the
contribution from triple scattering events in the \gpTHRj final state:
(a) a \gpONEj event mixed with a double-dijet DP event, where one jet 
of each dijet is lost or not reconstructed (Type I); 
(b) a DP event in the (\gpONEj)+dijet final state with one jet 
from the dijet lost, mixed with a dijet event with one jet lost (Type II).
Dashed lines correspond to the lost jets.}
\label{fig:tp_topol}
\end{figure}

In our previous study of DP \gpTHRj events
\cite{D0_2010}, we built a data-driven model of inclusive DP interactions (MixDP)
by combining \gpj and dijet events from data,
and obtaining the \gpTHRjX final state.
However, since each component of the MixDP model may contain
two (or more) jets, where one jet is due to an additional parton interaction, the model
simulates the properties of ``double plus triple'' parton interactions.
Therefore, the ``DP'' fractions found earlier in the \gpTHRj data (shown in Table III of \cite{D0_2010}) 
take into account a contribution from TP interactions as well.
These fractions are also shown in the second column of Table \ref{tab:frac_tab}.
Thus, if we calculate the fractions of TP events in the MixDP sample, defined as $f_{\rm tp}^{\rm dp+tp}$,
we can calculate the TP fractions in the \gpTHRj data, $f_{\rm tp}^{\gamma 3j}$, as 
\begin{equation}
 f_{\rm tp}^{\gamma 3j} = f_{\rm tp}^{\rm dp+tp} \cdot f_{\rm dp+tp}^{\gamma 3j},
\label{eq:tp_g3j}
\end{equation}
where $f_{\rm dp+tp}^{\gamma 3j}$ is the fraction of DP+TP events in the \gpTHRj sample.
Figure~\ref{fig:tp_topol} shows two possible ways in which a DP event and an SP event can be
combined to form a \gpTHRj event which is a part of the MixDP sample, with details on the origin
of the various parts of the event given in the caption.
Contributions from other possible MixDP configurations 
are negligible ($\lesssim 1\%$). 
In \cite{D0_2010}, we calculated how often each component, Type I and II, is found in the model. 
Table \ref{tab:frac_tab} shows that the events of Type II are dominant in all bins.
\begin{table}[htpb]
\small
\caption{Fractions of DP+TP events with total uncertainties in \gpTHRj data ($f_{\rm dp+tp}^{\gamma 3j}$)
and fractions of Type I~(II) events in the data-driven DP model ($F_{\rm Type~I(II)}$)
in the three $\ptsj$ bins.}
\label{tab:frac_tab}
\begin{tabular}{cccc} \hline\hline
$\ptsj$  & ~~$f_{\rm dp+tp}^{\gamma 3j}$ & $F_{\rm Type~I}$  & $F_{\rm Type~II}$  \\
 (GeV)   & ~~(\%)                        &        &                     \\\hline
$15-20$  & ~~$46.6\pm 4.1$               &  0.26  &  0.73      \\
$20-25$  & ~~$33.4\pm 2.3$               &  0.22  &  0.78      \\
$25-30$  & ~~$23.5\pm 2.7$               &  0.14  &  0.86      \\\hline\hline   
\end{tabular}
\end{table}
Thus, the fraction of TP configurations (Fig.~\ref{fig:tp_topol}) in the MixDP model, 
$f_{\rm tp}^{\rm dp+tp}$, can be calculated as
\begin{equation}
 f_{\rm tp}^{\rm dp+tp} = F_{\rm Type~II}\cdot f_{\rm dp}^{\gamma 2j} + F_{\rm Type~I}\cdot f_{\rm dp}^{jj},
\label{eq:tp_tpdp}
\end{equation}
where $f_{\rm dp}^{\gamma 2j}$ and $f_{\rm dp}^{jj}$ are 
the fractions of events with DP scattering resulting in \gpTWOj and dijet final states.
We separately analyze each of the event types of Fig.~\ref{fig:tp_topol}.
The  fraction of events having a second parton interaction with a dijet final state with cross section $\sigma^{jj}$
can be defined using the effective cross section $\sigma_{\rm eff}$ as $f_{\rm dp}^{jj} = \sigma^{jj}/(2\sigma_{\rm eff})$.
The cross section for a DP scattering producing two dijet final states can be presented then as
$\sigma_{\rm dp}^{jj,jj} = \sigma^{jj} f_{\rm dp}^{jj}$ \cite{TH3,Sjost}.
The fraction $f_{\rm dp}^{jj}$ is estimated using dijet events simulated with {\sc pythia}.
We calculate the jet cross sections $\sigma^{jj}$ for producing at least one jet
in the three $p_T$ bins with $|\eta^{\rm jet}|<3.5$.
The effective cross section $\sigma_{\rm eff}$ is taken as an average of the CDF \cite{CDF97} and 
D0 \cite{D0_2010} measurements, $\sigma_{\rm eff}^{\rm average} = 15.5$ mb.
The determined fractions are shown in the third column of Table \ref{tab:tp_frac_tot}.
We assume that the estimates, done at the particle level, are also approximately correct
at the reconstruction level. We take an uncertainty on these numbers 
$\delta f_{\rm dp}^{jj} = f_{\rm dp}^{jj}$.
\begin{table}[htpb]
\small
\caption{Fractions of DP events in \gpTWOj ($f_{\rm dp}^{\gamma 2j}$) and dijet ($f_{\rm dp}^{jj}$)
final states as well the fraction of TP configurations in the MixDP model ($f_{\rm tp}^{\rm dp+tp}$)
in the three $\ptsj$ bins.}
\label{tab:tp_frac_tot}
\begin{tabular}{cccc} \hline\hline
$\ptsj$ ~&~ $f_{\rm dp}^{\gamma 2j}$ & ~$f_{\rm dp}^{jj}$ & ~~~$f^{\rm dp+tp}_{\rm tp}$  \\
 (GeV)  ~&~ (\%)           ~&~  (\%)       ~&~ (\%)         \\\hline
$15-20$ ~&~  $15.9\pm 2.2$ ~&~ $0.50\pm 0.50$ ~&~ $11.7\pm 1.9$  \\
$20-25$ ~&~ ~$7.8 \pm 2.0$ ~&~ $0.17\pm 0.17$ ~&~ $ 6.1\pm 1.8$ \\
$25-30$ ~&~ ~$4.2 \pm 1.3$ ~&~ $0.07\pm 0.07$ ~&~ $ 3.6\pm 1.2$ \\\hline\hline  
\end{tabular}
\end{table}

The fractions of the \gpTWOj events in which the second jet is due to an additional
parton scattering are estimated in the previous section
and are much higher than $f_{\rm dp}^{jj}$. 
However, since we estimate the TP fraction in data at the reconstruction level, 
we repeat the same fitting procedure used for the extraction of $f_{\rm dp}^{\gamma 2j}$ from the
\dPhi distributions in the reconstructed data and SP \gpTWOj MC events.
The results of the fit in the three $\ptsj$ intervals 
are summarized in the second column of Table \ref{tab:tp_frac_tot}. 
Here the total uncertainties $\delta_{\rm tot}$ are due to the statistical and systematic uncertainties shown in Tables
\ref{tab:xs_dphi1} -- \ref{tab:xs_dphi3}, but excluding the uncertainties from the unfolding.

By substituting $f_{\rm dp}^{jj}$ and $f_{\rm dp}^{\gamma 2j}$ into Eq.~(\ref{eq:tp_tpdp}), we
calculate the TP fractions $f_{\rm tp}^{\rm dp+tp}$ in the MixDP model. 
They are shown in the last column of Table \ref{tab:tp_frac_tot}.
The TP fraction in the similar data-driven MixDP model 
in the CDF analysis \cite{CDF97} for (JES uncorrected) $5<p_T^{\rm jet2}<7$ GeV 
was estimated as $17^{+4}_{-8}\%$,
i.e., a value that is higher, on average, than our TP fractions measured 
at higher jet $p_T$, but in agreement with an extrapolation of our observed trend to lower jet $p_T$.

\begin{figure}[htbp]
\hspace*{0.0cm} \includegraphics[scale=0.37]{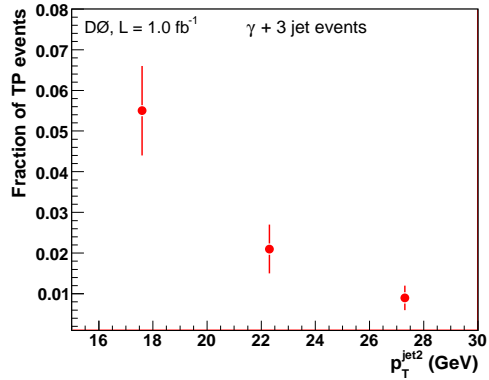}
\vskip -3mm
\caption{Fractions of TP events with total uncertainties in \gpTHRj final state as a function of $\ptsj$.}
\label{fig:tpfrac_ptj2}
\end{figure}

By substituting $f_{\rm tp}^{\rm dp+tp}$ and  the DP+TP fractions in \gpTHRj data $f_{\rm dp+tp}^{\gamma 3j}$ from \cite{D0_2010}
into Eq.~(\ref{eq:tp_g3j}), we get the TP fractions in the \gpTHRj data, $f_{\rm tp}^{\gamma 3j}$,
which are shown in the second column of Table \ref{tab:tp_frac}.
They are also presented in Fig.~\ref{fig:tpfrac_ptj2}.
The pure DP fractions, $f_{\rm dp}^{\gamma 3j}$, can then be obtained by subtracting
the TP fractions $f_{\rm tp}^{\gamma 3j}$ from the inclusive DP+TP fractions $f_{\rm dp+tp}^{\gamma 3j}$. 

The last column of Table \ref{tab:tp_frac} shows the ratios of the TP to DP fractions $f_{\rm tp}^{\gamma 3j}/f_{\rm dp}^{\gamma 3j}$ 
in \gpTHRj events. Since the probability of producing each additional parton scattering with a dijet final state
is expected to be directly proportional to $\sigma^{jj}/\sigma_{\rm eff}$, 
the $f_{\rm tp}^{\gamma 3j}/f_{\rm dp}^{\gamma 3j}$ ratio should be approximately
proportional to the jet cross section $\sigma^{jj}$, and drop correspondingly as a function of the jet $p_T$. 
This trend is confirmed in Table \ref{tab:tp_frac}. 
\begin{table}[htpb]
\small
\caption{Fractions of TP events (\%) and the ratio of TP/DP fractions in the three $\ptsj$ bins of \gpTHRj events. }
\label{tab:tp_frac}
\begin{tabular}{ccc} \hline\hline
$\ptsj$ ~&~ $f_{\rm tp}^{\gamma 3j}$ ~&~ $f_{\rm tp}^{\gamma 3j}/f_{\rm dp}^{\gamma 3j}$ \\
 (GeV)   ~&~ (\%)          &  \\\hline
$15-20$ ~&~ $5.5 \pm 1.1$ ~&~  ~$0.135 \pm 0.028$ \\
$20-25$ ~&~ $2.1 \pm 0.6$ ~&~  ~$0.066 \pm 0.020$ \\
$25-30$ ~&~ $0.9 \pm 0.3$ ~&~  ~$0.038 \pm 0.014$ \\\hline\hline
\end{tabular}
\end{table}

\section{Summary}
\label{Sec:conclusion}

We have studied the azimuthal correlations 
in \gpTHRj and \gpTWOj events and measured the normalized differential cross sections \DSigDS and \DSigDPhi
in three bins of the second jet $p_T$. The results are compared to different MPI models
and demonstrate that the predictions of the SP models do not describe the measurements
and an additional contribution from DP events is required to describe the data.
The data favor the predictions of the new {\sc pythia} MPI models with $p_T$-ordered showers, 
implemented in the Perugia and S0 tunes,
and also {\sc sherpa} with its default MPI model, while
predictions from previous {\sc pythia} MPI models, with tunes A and DW, are disfavored.

We have also estimated the fractions of DP events in the \gpTWOj samples
and found that they decrease in the bins of $\ptsj$ as 
$(11.6\pm1.0)\%$ for $15-20$ GeV, $(5.0\pm1.2)\%$ for $20-25$ GeV, 
and $(2.2 \pm 0.8)\%$ for $25-30$ GeV.
Finally, for the first time, we have estimated the fractions of TP events in the \gpTHRj data.
They vary in the $\ptsj$ bins as $(5.5\pm1.1)\%$ for $15-20$ GeV, $(2.1\pm0.6)\%$ 
for $20-25$ GeV, and $(0.9 \pm 0.3)\%$ for $25-30$ GeV.

The measurements presented in this paper can be used to improve the MPI models 
and reduce the existing theoretical ambiguities.
This is especially important for studies in which a dependence on MPI models is
a significant uncertainty (such as the top quark mass measurement),
and in searches for rare processes, for which DP events can be a sizable background.
~\\[3mm]
\centerline{\bf Acknowledgments}
%
We thank the staffs at Fermilab and collaborating institutions,
and acknowledge support from the
DOE and NSF (USA);
CEA and CNRS/IN2P3 (France);
FASI, Rosatom and RFBR (Russia);
CNPq, FAPERJ, FAPESP and FUNDUNESP (Brazil);
DAE and DST (India);
Colciencias (Colombia);
CONACyT (Mexico);
KRF and KOSEF (Korea);
CONICET and UBACyT (Argentina);
FOM (The Netherlands);
STFC and the Royal Society (United Kingdom);
MSMT and GACR (Czech Republic);
CRC Program and NSERC (Canada);
BMBF and DFG (Germany);
SFI (Ireland);
The Swedish Research Council (Sweden);
and
CAS and CNSF (China).

\section{Appendix}

In this appendix we discuss the unfolding procedure used to correct
the measured $\Delta S$ and $\Delta \phi$ distributions to the particle level
in order to obtain a nonparametric estimate of the true $\Delta S$ and $\Delta \phi$ distributions 
from the measured (reconstructed) distribution
taking into account possible biases and statistical uncertainties.
We use the following approach to extract the desired distributions from our measurements.
The observed distribution is the result of the convolution of a resolution function with the desired distribution
at the particle level.
After the discretization in $\Delta S$/$\Delta \phi$ bins, 
the resolution function is a smearing matrix, and distributions on both particle level
and reconstruction level become discrete distributions (i.e., histograms). 
The smearing matrix is stochastic: all elements are non-negative 
and the sum of elements in each column is equal to 1. Thus, the matrix columns are the probability density 
functions  that relate each bin in a histogram at the particle level to the bins in a histogram at the 
reconstruction level.
We split  the full $\Delta S$ ($\Delta \phi$) range $[0,\pi]$ into eight bins and
fix two bins at small $\Delta S$ ($\Delta \phi$) angles as $0-0.7$ and $0.7-1.2$ radians. 
These two bins are the most sensitive to a contribution from DP scattering, and 
their widths are chosen as a compromise between sensitivity to DP events and
the size of the relative statistical uncertainty.
The sizes of the other six bins are varied to minimize the ratio of maximum and minimum 
eigenvalues of the smearing matrix, defined as the condition number. 
This ratio, greater than unity, represents the scaling factor for 
the statistical uncertainties arising from the transformation of the differential cross sections
from the reconstructed to the particle level.
The same binning is used for the reconstructed and particle level distributions.
We build the smearing matrix using reconstructed and particle level events in the reweighted MC sample,
described in Sec.~\ref{Sec:Unf}.
To decrease the statistical uncertainties ($\delta_{\rm stat}$),
we use a Tikhonov regularization procedure \cite{Tikhonov,Anikeev,Anikeev1,Cowan} for the matrix.
This unfolding procedure may introduce a bias ($b$).
We optimize the regularization by finding a balance
between $\delta_{\rm stat}$ and $b$ according to the following criterion: 
we minimize the following function of $\delta_{\rm stat}$ and $b$ in the first two bins,
$0-0.7$ and $0.7-1.2$:
\vskip-4mm
\begin{equation}
U = \left[\left({0.5 b_1}\right)^2 + \left(\delta_{\rm stat,1}\right)^2 \right] +
\left[\left({0.5 b_2}\right)^2 + \left(\delta_{\rm stat,2}\right)^2 \right]. 
\label{eq:unf}
\end{equation}
\vskip -0mm
These two bins, being the most sensitive to contributions from DP scatterings, are 
the most important for our analysis.
We perform the regularization of the smearing matrix by adding a non-negative parameter $\alpha$ to all 
diagonal elements of the smearing matrix. 
The matrix columns are then re-normalized to make the matrix stochastic again
($\alpha=0$ is equivalent no regularization, while for $\alpha\to\infty$ 
the smearing matrix becomes the identity matrix).
The smallest uncertainties $U$ are usually achieved with $\alpha = 0.3-0.5$. 
An estimate of the unfolded distribution is obtained by multiplying the histogram (vector) of the measured  
$\Delta S$ and $\Delta \phi$ distributions by the inverted regularized smearing matrix. 
We use the sample of the reweighted  MC events 
to get an estimate of the statistical uncertainties and the bias in bins of the unfolded distribution. 
To accomplish this, we choose a MC subsample with the number of events equal to that of the selected 
data sample and having a discrete distribution (histogram) at the reconstruction level that is almost identical to that in data.
We randomize the MC histogram at the reconstruction level repeatedly (100,000 times)
 according to a multinomial distribution
and multiply this histogram by the inverted regularized smearing matrix for each perturbation. We obtain
a set of unfolded distributions at the particle level. 
Using this set and the true distribution for this MC sample, we estimate the statistical uncertainty
$\delta_{\rm stat}$ and the bias $b$ as the RMS and the mean of the distribution ``(true$-$unfolded)/true''
for each $\Delta S$ ($\Delta \phi$) bin.
The unfolded distribution is then corrected for the bias in each bin.
We assign half of the bias as a systematic uncertainty on this correction.
The overall unfolding corrections vary up to 60\%, being largest at the small angles.
The total uncertainties, estimated in each bin $i$ 
for the $\Delta S$ and $\Delta \phi$ distributions as  
$\sqrt{\left({0.5 b_i}\right)^2 + \left(\delta_{\rm stat,i}\right)^2}$,
vary between $10\%$ and  $18\%$.


\begin{thebibliography}{99}
\expandafter\ifx\csname natexlab\endcsname\relax\def\natexlab#1{#1}\fi
\expandafter\ifx\csname bibnamefont\endcsname\relax
  \def\bibnamefont#1{#1}\fi
\expandafter\ifx\csname bibfnamefont\endcsname\relax
  \def\bibfnamefont#1{#1}\fi
\expandafter\ifx\csname citenamefont\endcsname\relax
  \def\citenamefont#1{#1}\fi
\expandafter\ifx\csname url\endcsname\relax
  \def\url#1{\texttt{#1}}\fi
\expandafter\ifx\csname urlprefix\endcsname\relax\def\urlprefix{URL }\fi
\providecommand{\bibinfo}[2]{#2}
\providecommand{\eprint}[2][]{\url{#2}}

\bibitem{Landsh}
P.V.~Landshoff and J.C.~Polkinghorne, Phys. Rev. D {\bf 18}, 3344 (1978).

\bibitem{Goebel}
C.~Goebel, F.~Halzen, and D.M.~Scott, Phys. Rev. D {\bf 22}, 2789 (1980).

\bibitem{TH1}
F.~Takagi, Phys. Rev. Lett. {\bf 43}, 1296 (1979).

\bibitem{TH11}
N.~Paver and D.~Treleani, Nuovo Cimento A {\bf 70}, 215 (1982).

\bibitem{TH2}
B.~Humpert, Phys. Lett. B {\bf 131}, 461 (1983).

\bibitem{TH21}
B.~Humpert and R.~Odorico, Phys. Lett. B {\bf 154}, 211 (1985).

\bibitem{TH3}
T.~Sj$\ddot{\text o}$strand and M. van Zijl, Phys. Rev. D {\bf 36}, 2019 (1987).

\bibitem{Mang}
M.L.~Mangano, Z. Phys. C {\bf 42}, 331 (1989).

\bibitem{Sjost}
T.~Sj$\ddot{\text o}$strand and P.Z.~Skands,  J. High Energy Physics {\bf 0403}  (2004).

\bibitem{Perugia}
P.Z.~Skands, 
Fermilab-CONF-09-113-T.

\bibitem{PYT}

T.~Sj\"ostrand {\sl et~al.}, J. High Energy Physics {\bf 0005}, 026 (2006).

\bibitem{pT_order}
T.~Sj\"ostrand and P.Z.~Skands, Eur. Phys. J. C {\bf 39}, 129 (2005). 

\bibitem{Snigir}
A.M.~Snigirev, Phys. Rev. D {\bf 68}, 114012 (2003).

\bibitem{Snigir1}
V.L.~Korotkikh and A.M.~Snigirev, Phys. Lett. B {\bf 594}, 171 (2004).

\bibitem{Trel_05}
E.~Cattaruzza, A.~Del Fabbro, and D.~Treleani, Phys. Rev. D {\bf 72}, 034022 (2005).

\bibitem{GS}
J.R.~Gaunt and W.J.~Stirling, J. High Energy Physics {\bf 1003}, 005  (2010). 

\bibitem{Snigir2}
A.M.~Snigirev, Phys. Rev. D {\bf 81}, 065014  (2010). 

\bibitem{AFS} 
T.~Akesson {\sl et al.} (AFS Collaboration), Z. Phys. C {\bf 34}, 163 (1987). 


\bibitem{UA2} 
J.~Alitti {\sl et al.} (UA2 Collaboration), Phys. Lett. B {\bf 268}, 145 (1991).

\bibitem{CDF93}
F.~Abe {\sl et al.} (CDF Collaboration), Phys. Rev. D {\bf 47}, 4857 (1993).

\bibitem{CDF97} 
F.~Abe {\sl et al.} (CDF Collaboration), Phys. Rev. D {\bf 56}, 3811 (1997).

\bibitem{E735}
T.~Alexopoulos {\sl et al.} (E735 Collaboration), Phys. Lett. B {\bf 435}, 453 (1991).

\bibitem{D003} 
V.M.~Abazov {\sl et al.} (D0 Collaboration), Phys. Rev. D {\bf 67}, 052001 (2003).

\bibitem{D0_2010} 
V.M.~Abazov {\sl et al.} (D0 Collaboration), Phys. Rev. D {\bf 81}, 052012 (2010).

\bibitem{ZEUS} 
C.~Gwenlan {\sl et al.} (ZEUS Collaboration), Acta Phys. Polon. B {\bf 33}, 3123 (2002).

\bibitem{H1} 
A.~Knutsson (H1 and ZEUS Collaborations), Nucl. Phys. Proc. Suppl. {\bf 191}, 141 (2009).

\bibitem{WH} 
D.~Fabbro and D.~Treleani, Phys. Rev. D {\bf 61}, 077502 (2000).

\bibitem{WH1} 
D.~Fabbro and D.~Treleani,
Phys. Rev. D {\bf 66}, 074012 (2002).

\bibitem{Huss} 
M.Y.~Hussein, Nucl. Phys. Proc. Suppl. {\bf 174}, 55 (2007).  

\bibitem{Berger_dp} 
E.L.~Berger, C.B.~Jackson and G.~Shaughnessy, Phys. Rev. D {\bf 81}, 014014 (2010). 

\bibitem{DPWH}
D.V.~Bandurin, G.A.~Golovanov and N.B.~Skachkov, arXiv:1011.2186 [hep-ph] (2010).

\bibitem{Skands_top} 
D.~Wicke and P.Z.~Skands, Nuovo Cimento B {\bf 123} (2008). 

\bibitem{CDF_gj_angles} 
F.~Abe {\sl et al.} (CDF Collaboration), Phys. Rev. D {\bf 57}, 67 (1998).


\bibitem{ToolsAndJets}
C.~Buttar {\sl et~al.}, arXiv:0803.0678 [hep-ph].


\bibitem{Fig9}
See Fig. 9 of \cite{D0_2010} and nearby text.

\bibitem{D0_det}
V.M.~Abazov {\sl et al.} (D0 Collaboration), Nucl. Instrum. Methods Phys. Res. A {\bf 565}, 463 (2006).

\bibitem{etaphi}
The polar angle $\theta$ and the azimuthal angle $\phi$ are defined
with respect to the positive $z$ axis, which is along the proton beam direction.
Pseudorapidity is defined as $\eta=-\ln[\tan(\theta/2)]$.
$\eta_{\text {det}}$ and $\phi_{\text {det}}$ are the pseudorapidity and the azimuthal angle
measured with respect to the center of the detector.

\bibitem{gamjet_xs}
V.M.~Abazov {\sl et al.} (D0 Collaboration), 
Phys. Lett. B {\bf 666}, 2435 (2008).

\bibitem{Run2Cone}
G.C.~Blazey {\sl et~al.}, arXiv:hep-ex/0005012 (2000).



\bibitem{Sherpa} 
T.~Gleisberg {\sl et~al.}, J. High Energy Physics {\bf 0902}, 007 (2009). 

\bibitem{CKKW} 
S.~Catani {\sl et al.}, 
J. High Energy Physics {\bf 0111}, 063 (2001). 

\bibitem{Sherpa_photons} 
S.~H\"oche, S.~Schumann, and F.~Siegert, Phys. Rev. D {\bf 81}, 034026 (2010). 

\bibitem{Sherpa_matching} 
The choice of ME-PS matching parameters was discussed with the {\sc sherpa} authors.

\bibitem{JN} 
C.~Peterson, T.~Rognvaldsson, and L.~L{\"o}nnblad, Comput. Phys. Commun. {\bf 81}, 15 (1994).

\bibitem{HMCMLL} 
R.~Barlow and C.~Beeston,  Comput. Phys. Commun. {\bf 77}, 219 (1993). 

\bibitem{GEANT}
R.~Brun and F.~Carminati, CERN Program Library Long Writeup W5013 (1993).

\bibitem{Tikhonov}
A.N.~Tikhonov, A.S.~Leonov, and A.G.~Yagola, ``Nonlinear ill-posed problems'', Vols. 1, 2 (Chapman and Hall, London, 1998).

\bibitem{Anikeev}
V.B.~Anykeev, A.A.~Spiridonov, and V.P.~Zhigunov, 
Nucl. Instrum. Methods Phys. Res. A {\bf 303},  350 (1991).

\bibitem{Anikeev1}
V.B.~Anikeev and V.P.~Zhigunov, Phys. Part. Nucl. {\bf 24(4)}, 424 (1993).

\bibitem{Cowan}
G.~Cowan, ``Statistical Data Analysis'' (Oxford University Press, 1998).

\bibitem{JETPHOX}
P.~Aurenche {\sl et al.}, {\sc jetphox} package, \\
http://wwwlapp.in2p3.fr/lapth/{\sc phox}$_-${\sc family}/main.html.



\end{thebibliography}
\end{document}